\documentclass[10pt, conference, letterpaper]{IEEEtran}
\IEEEoverridecommandlockouts
\usepackage{cite}
\usepackage{amsmath,amssymb,amsfonts}
\usepackage[ruled,noend,linesnumbered]{algorithm2e} %linesnumbered,

\usepackage{algpseudocode}
\usepackage{graphicx}
\usepackage{xcolor}% http://ctan.org/pkg/xcolor
\usepackage{graphicx}
\usepackage{textcomp}
\usepackage{xcolor}
\usepackage{enumitem}
\usepackage{url}
\definecolor{red}{rgb}{1,0,0} %Black:{0,0,0} Red:{1,0,0}
\definecolor{blue}{rgb}{0,0,1} %Black:{0,0,0} Red:{1,0,0}
\definecolor{dispatchers1}{rgb}{0.412,0.392,0.482} %Black:{0,0,0} Red:{1,0,0}
\definecolor{orchestrator1}{rgb}{0.69,0.478,0.596} %Black:{0,0,0} Red:{1,0,0}
\definecolor{dispatchers2}{rgb}{0.380,0.227,0.545} %Black:{0,0,0} Red:{1,0,0}
\definecolor{orchestrator2}{rgb}{0.525,0.220,0.271} %Black:{0,0,0} Red:{1,0,0}

\def\BibTeX{{\rm B\kern-.05em{\sc i\kern-.025em b}\kern-.08em
    T\kern-.1667em\lower.7ex\hbox{E}\kern-.125emX}}

\begin{document}
\title{Tailored Learning-Based Scheduling for \textit{Kubernetes}-Oriented Edge-Cloud System}
\author{
    \IEEEauthorblockN{Yiwen Han\IEEEauthorrefmark{2}, Shihao Shen\IEEEauthorrefmark{2}, Xiaofei Wang\IEEEauthorrefmark{2}, Shiqiang Wang\IEEEauthorrefmark{5},
		Victor C.M. Leung\IEEEauthorrefmark{3}\IEEEauthorrefmark{4}}
    \IEEEauthorblockA{\IEEEauthorrefmark{2}TANKLab, College of Intelligence and Computing, Tianjin University, Tianjin, China}
    \IEEEauthorblockA{\IEEEauthorrefmark{5}IBM T. J. Watson Research Center, Yorktown Heights, NY, USA}
    \IEEEauthorblockA{\IEEEauthorrefmark{3}College of Computer Science and Software Engineering, Shenzhen University, Shenzhen, China}
    \IEEEauthorblockA{\IEEEauthorrefmark{4}Dept. of Electrical and Computer Engineering, the University of British Columbia, Vancouver, Canada}
	\IEEEauthorblockA{\{hanyiwen, shenshihao, xiaofeiwang\}@tju.edu.cn, shiqiang.wang@ieee.org, vleung@ieee.org}
	\vspace{-1.5em}
    \thanks{The code of \textit{KaiS} is available at:
	\textbf{\textit{\protect\url{https://github.com/XiaofeiTJU/KaiS}}}

Xiaofei Wang was supported by the National Science Foundation of China (No. 62072332) and the National Key R\&D Program of China (No.~2019YFB2101901 and No.~2018YFC0809803); Victor C. M. Leung was supported by Chinese National Engineering Laboratory for Big Data System Computing Technology and Canadian Natural Sciences and Engineering Research Council. Corresponding author: \textit{Xiaofei Wang}. }
}

%\title{Conference Paper Title*\\

%\and
%\and
%\and
%\and
%\and
\maketitle
\begin{abstract}
\textit{Kubernetes} (\textit{k8s}) has the potential to merge the distributed edge and the cloud but lacks a scheduling framework specifically for edge-cloud systems.
Besides, the hierarchical distribution of heterogeneous resources and the complex dependencies among requests and resources make the modeling and scheduling of \textit{k8s}-oriented edge-cloud systems particularly sophisticated.
In this paper, we introduce \textit{KaiS}, a learning-based scheduling framework for such edge-cloud systems to improve the long-term throughput rate of request processing.
First, we design a coordinated multi-agent actor-critic algorithm to cater to decentralized request dispatch and dynamic dispatch spaces within the edge cluster.
Second, for diverse system scales and structures, we use graph neural networks to embed system state information, and combine the embedding results with multiple policy networks to reduce the orchestration dimensionality by stepwise scheduling.
Finally, we adopt a two-time-scale scheduling mechanism to harmonize request dispatch and service orchestration, and present the implementation design of deploying the above algorithms compatible with native \textit{k8s} components.
Experiments using real workload traces show that \textit{KaiS} can successfully learn appropriate scheduling policies, irrespective of request arrival patterns and system scales.
Moreover, \textit{KaiS} can enhance the average system throughput rate by $14.3\%$ while reducing scheduling cost by $34.7\%$ compared to baselines.
%The computing paradigm is gradually shifting to end-edge-cloud computing to integrate and exploit the potential of computing resources emerging at the network edge.
%However, the hierarchical distribution of heterogeneous resources and the complex dependencies among requests and resources make the modeling and scheduling of the edge-cloud system particularly sophisticated.
%In this paper, we introduce \textit{KaiS}, a learning-based scheduling framework for \textit{kubernetes (k8s)}-oriented edge-cloud systems to improve the long-term throughput of request processing.
%First, we design a coordinated multi-agent actor-critic algorithm to cater to decentralized request dispatch and dynamic dispatch spaces within the edge cluster.
%Second, for diverse system scales and structures, we use graph neural networks to embed system state information, and combine the embedding results with multiple policy networks to reduce orchestration dimensionality by stepwise scheduling.
%Finally, we adopt a two-time-scale scheduling mechanism to harmonize request dispatch and service orchestration, and present the implementation design of deploying the above algorithms compatible with native \textit{k8s} components.
%Experiments using real workload traces show that \textit{KaiS} can successfully learn appropriate scheduling policies, irrespective of service request patterns and system scales.
%Moreover, \textit{KaiS} can enhance the average system throughput rate by $14.3\%$ while reducing scheduling cost by $34.7\%$ compared to baselines.
\end{abstract}

%\begin{IEEEkeywords}
%End-Edge-Cloud Computing, Edge Computing, Kubernetes, Reinforcement Learning
%\end{IEEEkeywords}
\vspace{-0.4em}
\section{Introduction}
\label{sec:Introduction}

\vspace{-0.3em}
\subsection{Background and Problem Statement}
\vspace{-0.3em}

To provide agile service responses and alleviate backbone networks, edge and cloud computing are gradually converging to achieve this goal by hosting services as close as possible to where requests are generated \cite{Shi2016, Ren2019b}.
Edge-cloud systems are commonly built on \textit{Kubernetes} (\textit{k8s}) \cite{Burns2016,kubeedge,openyurt,baetyl} and are designed to seamlessly integrate distributed and hierarchical computing resources at the edge and the cloud \cite{Wang2020}.
One fundamental problem for supporting efficient edge-cloud systems is: how to schedule request dispatch \cite{Tan2017} and service orchestration (placement) \cite{Pasteris2019} within the \textit{k8s} architecture.
However, native \textit{k8s} architecture is hard to manage the geographically distributed edge computing resources, while the customized edge-cloud frameworks (e.g., \textit{KubeEdge} \cite{kubeedge}, \textit{OpenYurt} \cite{openyurt} and \textit{Baetyl} \cite{baetyl}) based on \textit{k8s} do not address the above scheduling issues.

\begin{figure}[!!!!!!!!!!!!!!hhhhhhhhhht]%[!htp]
    \centering
%  \vspace{-0.8em}
    \includegraphics[width=8.85 cm]{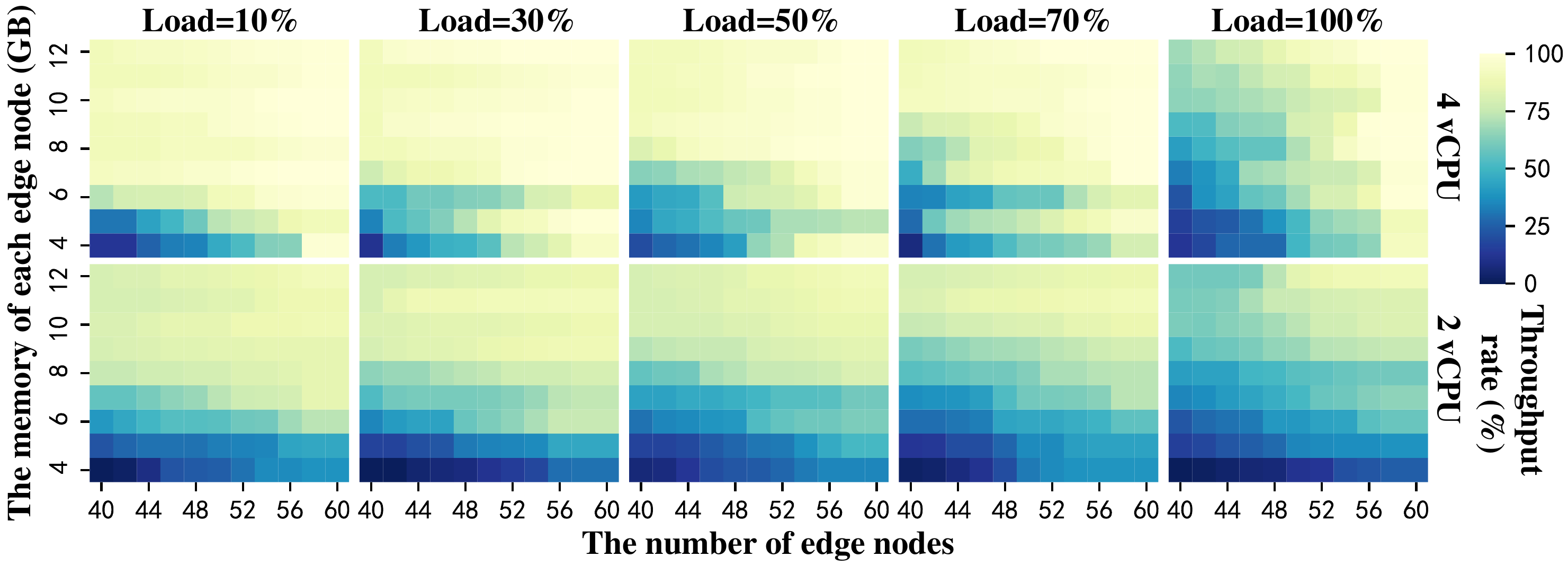}
\setlength{\abovecaptionskip}{-0.6cm} 
    \caption{The model of system throughput is highly complex and non-linear.}
  \vspace{-1.8em}
    \label{fig:NonlinearEffect}
\end{figure}

To serve various requests, the edge-cloud system needs to manage corresponding service entities across edge and cloud while able to determine where these requests should be processed. 
Though \textit{k8s} is the most popular tool for managing cloud-deployed services, it is not yet able to accommodate both edge and cloud infrastructure and support request dispatch at the distributed edge. 
In this case, how to (\textit{$\romannumeral1$}) adapt \textit{k8s} components and extend its current logic to bind the distributed edge and the cloud and (\textit{$\romannumeral2$}) devise scheduling algorithms that can fit in \textit{k8s} is the key for efficient edge-cloud systems.

\vspace{-0.4em}
\subsection{Limitations of Prior Art and Motivation}
\vspace{-0.2em}
Most scheduling solutions for request dispatch and service orchestration rely on the accurate modeling or prediction of service response times, network fluctuation, request arrival patterns, etc. \cite{Farhadi2019, Poularakis2019, Ma2020}. 
Nevertheless, (\textit{$\romannumeral1$}) \textit{the heterogeneous edge nodes and the cloud cluster are connected in uncertain network environments, and practically form a dynamic and hierarchical computing system}. 
As shown in Fig.~\ref{fig:NonlinearEffect}, the system behavior, i.e., the average throughput rate of that system managed by native \textit{k8s}, substantially varies with the available resources and the request loads (refer to Sec. \ref{sec:Performance Evaluation} for detailed settings).
More importantly, (\textit{$\romannumeral2$}) \textit{the underlying model that captures this behavior is highly nonlinear and far from trivial}.
However, even though rich historical data are available, it is hard to achieve the exact estimation of these metrics \cite{Wang2020, Ayala-Romero2019} and then design scheduling policies for any specific request arrivals, system scales and structures, or heterogeneous resources. 
Further, (\textit{$\romannumeral3$}) \textit{few solutions carefully consider whether the proposed scheduling framework or algorithms are applicable to the actual deployment environment}, i.e., whether they are compatible with \textit{k8s} or others to integrate with the existing cloud infrastructure.
Therefore, a scheduling framework for a \textit{k8s}-oriented edge-cloud system, without relying on the assumption about system dynamics, is desired.

\begin{figure}[t]%[!htp]
    \centering
%  \vspace{-0.6em}
    \includegraphics[width=8.85 cm]{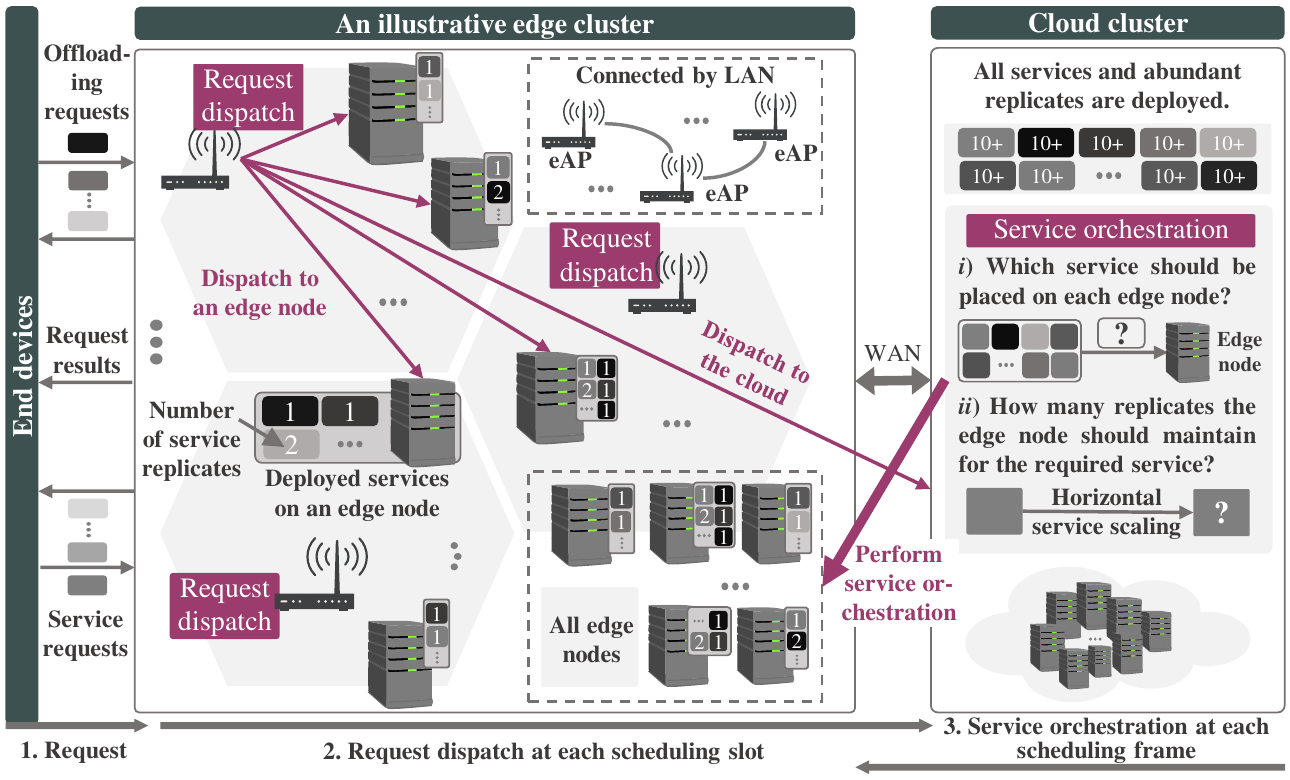}
\setlength{\abovecaptionskip}{-0.5cm} 
    \caption{Scheduling in \textit{kubernetes}-oriented edge-cloud system.}
  \vspace{-1.7em}
    \label{fig:end-edge-cloud computing}
\end{figure}

\vspace{-0.4em}
\subsection{Technical Challenges and Solutions}
\vspace{-0.2em}
In this paper, we show that learning techniques \cite{sutton2018reinforcement} can help edge-cloud systems by automatically learning effective system scheduling policies to cope with stochastic service request arrivals.
We propose \textit{KaiS}, a \textit{\d{k}8s}-oriented and le\d{a}rn\d{i}ng-based scheduling framework for edge-cloud \d{s}ystems.
Given only a high-level goal, e.g., to maximize the long-term throughput of service processing, \textit{KaiS} automatically learns sophisticated scheduling policies through experiences of the system operation, without relying on the assumptions about system execution parameters and operating states.
%\textit{KaiS} adopts two-time-scale scheduling inspired by \cite{Farhadi2019}.
%Specifically, at a smaller time scale (slot), it reasonably dispatches service requests to edge nodes or the cloud for processing according to the available resources. 
%At a larger scale (frame), it regularly orchestrates deployed service entities through releasing and capturing resources.
%\textit{KaiS} tailors Deep Reinforcement Learning (DRL) to learn scheduling policies through experiences of the system running, and encodes learned policies in neural networks, without relying on various inaccurate assumptions about system execution parameters and operating states.
%Our objective is not to apply off-the-shelf DRL algorithms, such as DQN \cite{Mnih2015} and DDPG \cite{lillicrap2015continuous}, to scheduling problems.
To guide \textit{KaiS} to learn scheduling policies, we need to tailor learning algorithms in the following aspects: \textit{the coordinated learning of multiple agents, the effective encoding of system states, the dimensionality reduction of scheduling actions, etc.}

For request dispatch, as depicted in Fig.~\ref{fig:end-edge-cloud computing}, \textit{KaiS} needs to scale to hundreds of distributed edge Access Points (eAPs) \cite{Ren2019b}.
Traditional learning algorithms, such as DQN \cite{Mnih2015} and DDPG \cite{lillicrap2015continuous}, that usually use one centralized learning agent, is not feasible for \textit{KaiS} since the distributed eAPs will incur dispatch action space explosion \cite{WangMADRL}.
To ensure timely dispatch, \textit{KaiS} requires the dispatch action be determined at where the request arrives, i.e., eAPs, in a decentralized (instead of centralized) manner \cite{WangMADRL}.
Thus, we leverage Multi-Agent Deep Reinforcement Learning (MADRL) \cite{MARLSurvey} and place a dispatch agent at each eAP.
%However, such settings (\textit{$\romannumeral1$}) require numerous agents to interact with the system at each time and (\textit{$\romannumeral2$}) result in that their action spaces vary with available system resources, making them difficult to learn scheduling policies.
However, such settings (\textit{$\romannumeral1$}) require numerous agents to interact with the system at each time and (\textit{$\romannumeral2$}) have varying dispatch action spaces that depend on available system resources, making these agents difficult to learn scheduling policies.
Hence, we decouple centralized critic and distributed actors, feeding in global observations during critic training to stabilize each agent's learning process, and design a policy context filtering mechanism for actors to respond to the dynamic changes of dispatch action space.

Besides, \textit{KaiS} must orchestrate dozens of or more services according to the system's global resources and adapt to different system scales and structures.
Hence, \textit{KaiS} requires our learning techniques to (\textit{$\romannumeral1$}) encode massive and diverse system state information, and (\textit{$\romannumeral2$}) represent bigger and complex action space for orchestration.
Thus, we employ Graph Neural Networks (GNNs) \cite{Zhang2020} and multiple policy networks \cite{Silver2014} to encode the system information and reduce the orchestration dimensionality, respectively, without manual feature engineering.
Compared with common DRL solutions with raw states and fixed action spaces, our design can reduce model complexity, benefiting the learning of scheduling policies.
%%%%%%%%%%%%%%%%%%%%%%%%%%%%%%%%%%%%%%%%%%%%%%
%Finally, for both request dispatch and service orchestration, traditional DRL algorithms cannot cope with continuously arriving service requests.
%Due to the randomness of arriving requests, the learning algorithm cannot distinguish whether the impact of the two scheduling decisions should be attributed to different service request patterns or the quality of themselves.
%We condition training feedback \cite{mao2018variance} to associate the decision impact with the actual service request arrival sequence as much as possible and isolate the contributions of the scheduling policy in the overall feedback, making it feasible to learn scheduling policies that can deal with stochastic request arrivals.
%In addition, the learning-based scheduling policy is bound to make bad decisions in the early stage of training.
%Therefore, under the situation that service requests continue to arrive, a scheduling policy that has not been well-trained will inevitably reduce the system throughput, resulting in a backlog of a large number of service requests and untimely processing.
%In this case, continuing to spend a lot of training time to explore better actions cannot improve the scheduling policy \cite{CurriculumLearning}.
%In order to solve this issue, we first use simple and short service request sequences for training, and then moderately introduce more sophisticated request sequences step by step, so that the scheduling policy can be gradually improved.
\vspace{-0.8em}
\subsection{Main Contributions}
\vspace{-0.3em}
%In summary, our main contributions are as follows:
\begin{itemize}[leftmargin=*]
\item A coordinated multi-agent actor-critic algorithm for decentralized request dispatch with a policy context filtering mechanism that can deal with dynamic dispatch action spaces to address time-varying system resources.
\item A GNN-based policy gradient algorithm for service orchestration that employs GNNs to efficiently encode system information and multiple policy networks to reduce orchestration dimensionality by stepwise scheduling.
\item A two-time-scale scheduling framework implementation of the tailored learning algorithms for the \textit{k8s}-oriented edge-cloud system, i.e., \textit{KaiS}, and an evaluation of \textit{KaiS} with real workload traces in various scenarios and against baselines.
%\item A \textit{k3s}-oriented system design for edge-cloud collaboration, with a set of decentralized request dispatchers deployed at the edge and a centralized service orchestrator at the cloud, that can perform two-time-scale scheduling to handle stochastic arriving service requests.
%\item Learning techniques that can aware of the contextual dynamics of the system and learn scheduling policies without human input, including a multi-agent DRL algorithm designed to tackle request dispatch in a decentralized manner and a GNN-based DRL approach tailored for orchestrating services deployed at the edge. 
%\item A prototype implementation, i.e., \textit{KaiS}, of the learning-based scheduling algorithms for that \textit{k3s}-oriented system, and an evaluation with real-world workload traces in various scenarios and against other baselines. 
\end{itemize}

%%%%%%%%%%%%%%%%%%%%%%%%%%%%%
%To the best of our knowledge, this is the first paper in the literature that thoroughly explores empirically the learning-based scheduling behavior of the end-edge-cloud computing within a \textit{k3s}-based experimental setup. 
%Our results do not only shed light on the potential of combining \textit{k3s} components with request dispatchers and the service orchestrator, but also show that substantial gains can be achieved by tailoring learning algorithms that adapt to dynamic contexts in the proposed framework.

In the sequel, Sec. \ref{sec:Scheduling Problem Statement} introduces the scheduling problem. 
Sec. \ref{sec:Algorithm and System Design} and Sec. \ref{sec:System Design and Implementation} elaborate the algorithm and implementation design. 
Sec. \ref{sec:Performance Evaluation} presents experiment results.
Finally, Sec. \ref{sec:Related Work} reviews related works and Sec. \ref{sec:Conclusion} concludes the paper. 

%\section{Background}
%
%\subsection{End-Edge-Cloud Computing and Kubernetes}
%
%\subsection{Multi-agent Deep Reinforcement Learning}
%
%\subsection{Graph Neural Networks-based Reinforcement Learning}

\vspace{-0.4em}
\section{Scheduling Problem Statement}
\label{sec:Scheduling Problem Statement}
\vspace{-0.3em}

We focus on scheduling request dispatch and service orchestration for the edge-cloud system to \textit{improve its long-term throughput rate}, i.e., the ratio of processed requests that meet delay requirements during the long-term system operation.

\vspace{-0.8em}
\subsection{Edge-Cloud System}
\vspace{-0.3em}

As shown in Fig.~\ref{fig:end-edge-cloud computing}, neighboring eAPs and edge nodes form a resource pool, i.e., an edge cluster, and connect with the cloud.
When requests arrive at eAPs, the edge cluster then handles them together with the cloud cluster $C$.
For clarity, \textit{we only take one edge cluster to exemplify \textit{KaiS}}, and consider the case that there is no cooperation between geographically distributed edge clusters.
Nonetheless, by maintaining a service orchestrator for each edge cluster, \textit {KaiS} can be easily generalized to support geographically distributed edge clusters.
\begin{itemize}[leftmargin=*]
\item \textbf{\textit{Edge Cluster and Edge Nodes}}. To process requests, the edge cluster should host corresponding service entities. 
An edge cluster consists of a set $\mathcal{B} = \{ 1, 2, \ldots, B \}$ of eAPs indexed by $b$, and $\mathcal{N}_{b} = \{ 1, 2, \ldots, N_{b} \}$ is the set of edge nodes attached to and managed by eAP $b$.
All edge nodes in the edge cluster are represented by $\mathcal{N} = \{ 1, 2, \ldots, N \}$.
%, namely $\mathcal{N} = \{ \mathcal{N}_b: b \in \mathcal{B} \} $
All eAPs, along with associated edge nodes, are connected by Local Area Network (LAN).
A request arrived at the edge can be dispatched to an edge node or the cloud by the eAP that admits it for processing.
\item \textbf{\textit{Cloud Cluster}}. The cloud cluster has sufficient computing and storage resources compared to the edge and is connected to eAPs through WAN (Wide Area Network), 
It can undertake requests that edge clusters cannot process.
In addition, it manages all geographically distributed edge clusters, including orchestrating all service entities in each edge cluster according to the system's available resources.
\end{itemize}

%\subsubsection{End Devices}
%
%\subsubsection{Edge Cluster and Edge Nodes}
%
%\subsubsection{Cloud Cluster}

\vspace{-1em}
\subsection{Scheduling to Improve Long-term System Throughput}
\label{subsec:Scheduling in End-Edge-Cloud Computing}
\vspace{-0.25em}

We adopt a two-time-scale mechanism \cite{Farhadi2019} to schedule request dispatch and service orchestration, i.e., \textit{KaiS} performs request dispatch at a smaller scale, slot $t$, while carrying out service orchestration at a larger scale, frame $\tau = \beta t$ ($\beta \in \mathbb{N}^+$).

%\vspace{-0.15em}
\textbf{\textit{Dispatch of Requests at eAPs}}. 
Delay-sensitive service requests are stochastically arriving at eAPs.
For each eAP $b \in \mathcal{B}$, it maintains a queue $\mathcal{Q}_{b}$ for the requests arrived at it, and a dynamic dispatch policy $\hat{\pi}_{b, t}$ varying with time.
According to $\hat{\pi}_{b, t}$, at each slot $t$, each eAP $b$ dispatches a request to an edge node, where the required service entity is deployed and that has sufficient resources, or the cloud cluster with sufficient computing resources for processing.
The processing of each request consumes both computation resources and network bandwidth of the edge or the cloud. 
Moreover, dispatching requests to the cloud may lead to extra transmission delay since it is not as close to end devices, i.e., where requests are generated.
Each edge node and the cloud maintain a queue of dispatched requests, i.e., $\{ \mathcal{Q}_n: n \in \mathcal{N} \}$ for edge nodes and $\mathcal{Q}_{C}$ for the cloud, and process their respective queue by a specific strategy, e.g., prioritizing requests with strict delay requirements.
To ensure timely scheduling, it is ideal to have the eAPs, where requests first arrive, perform request dispatch independently, instead of letting the cloud or the edge to make dispatching decisions in a centralized manner, since it may incur high scheduling delays \cite{Chen2016f}.
For requests that are not processed in time, the system drops them at each slot.

\textbf{\textit{Orchestration of Services at Edge Cluster}}. Due to the storage capacity and memory limit of edge nodes, not all services $\mathcal{W} = \{ 1, \ldots, w, \ldots, W \}$ can be stored and hosted on each of them.
In this case, service entities at the edge cluster should be orchestrated, which includes the following questions: (\textit{$\romannumeral1$}) which service should be placed on which edge node and (\textit{$\romannumeral2$}) how many replicates the edge node should maintain for that service. 
Besides, service requests arrivals at different times may have different patterns, resulting in the intensity of demand for different services varying over time.
Hence, the scheduling should be able to capture and identify such patterns and, based on them, to orchestrate services to fulfill stochastically arriving requests.
Unlike request dispatch, too frequent large-scale service orchestration in the edge cluster may incur system instability and high operational costs \cite{Farhadi2019}.
A more appropriate solution is to have the cloud perform service orchestration for the edge with a dynamic scheduling policy $\tilde{\pi}_{\tau}$ at each frame $\tau$.
Based on $\tilde{\pi}_{\tau}$, the cloud determines $\tilde{d}_{w,n} \in \mathbb{N}$, i.e., the number of replicates of service $w$ on edge node $n$ during frame $\tau$.
Particularly, $\tilde{d}_{w,n} = 0$ means that edge node $n$ does not host service $w$.

%Our objective is to maximize the long-term system throughput.
%Hence, we introduce the \textbf{\textit{effective throughput rate per frame}} $\varPhi_r = \sum\nolimits_{b \in \mathcal{B}}\sum\nolimits_{n_b \in \mathcal{N}_b} \varUpsilon_{\tau}(\mathcal{Q}_{n_b}) / \bar{\varUpsilon}_{\tau}(\mathcal{Q}_{n_b})$, which is the ratio of requests, completed within delay requirements and in a frame, to the total backlogged and newly arrived requests before the end of that frame.
%In this case, our scheduling problem can be formulated as
The scheduling objective is to maximize the long-term system throughput $\varPhi = \sum\nolimits_{\tau}^{\infty}\sum\nolimits_{n \in \mathcal{N}} \varUpsilon_{\tau}(\mathcal{Q}_{n}) +\varUpsilon_{\tau}(\mathcal{Q}_{C})$, where $\varUpsilon_{\tau}(\mathcal{Q}_{n})$, $\varUpsilon_{\tau}(\mathcal{Q}_{C})$ represent the number of requests that have been processed timely by edge node $n$ or the cloud in frame $\tau$, respectively.
%However, since it is not possible to measure $\varPhi$, we turn to a more realistic metric, i.e., \textbf{\textit{throughput rate}} $\varPhi_r$, which is the ratio of requests, completed within delay requirements and in a frame, to the total backlogged and newly arrived requests before the end of that frame.
%If $\varPhi_r$ can be maximized at each frame, we can realize the goal of improving the long-term system throughput rate.
To avoid $\varPhi \rightarrow \infty$, we use a more realistic metric, i.e., the long-term system throughput rate $\varPhi^{\prime} \in [0,1]$, which is the ratio of requests, completed within delay requirements, to the total number of arrived requests at the system.
The long-term throughput rate $\varPhi^{\prime}$ can be denoted as $\varPhi^{\prime} = \varPhi / \sum\nolimits_{\tau}^{\infty}\sum\nolimits_{b \in \mathcal{B}} \bar{\varUpsilon}_{\tau}(\mathcal{Q}_{b}) $, where $\bar{\varUpsilon}_{\tau}(\mathcal{Q}_{b})$ indicates the number of requests arrived at eAP $b$ during frame $\tau$.
In this case, our scheduling problem for both request dispatch and service orchestration can be formulated as
\begin{equation}
\label{equ:OptimizationProblem}
\underset{  \{\hat{\pi}_{b, t} : b \in \mathcal{B} \}, \tilde{\pi}_{\tau} }{\max} \varPhi^{\prime} = \underset{  \{\hat{\pi}_{b, t} : b \in \mathcal{B} \}, \tilde{\pi}_{\tau} }{\max} \varPhi / \sum_{\tau}^{\infty}\sum_{b \in \mathcal{B}} \bar{\varUpsilon}_{\tau}(\mathcal{Q}_{b}) ,
\end{equation}
where, for clarity, we use scheduling policies $\{ \hat{\pi}_{b, t}:b \in \mathcal{B} \}$ and $\tilde{\pi}_{\tau}$ instead of a series of scheduling variables at slots and frames to represent the problem.
Compared to the problem in \cite{Farhadi2019}, our scheduling is more complicated since it involves integer dispatch variables.
More details on the constraints and NP-hard proof of such a long-term scheduling problem can be found in \cite{Farhadi2019}.
In this work, we tailor learning algorithms for \textit{KaiS} to improve the long-term system throughput rate.

%\begin{equation}
%\label{equ:OptimizationProblem}
%\underset{  \{\hat{\pi}_{b, t}, b \in \mathcal{B} \}, \tilde{\pi}_{\tau = \beta t} }{\max} \sum_{t}^{\infty}\sum_{b \in \mathcal{B}}\sum_{n_b \in \mathcal{N}_b} \varUpsilon_{t}(\mathcal{Q}_{n_b}) / \bar{\varUpsilon}_{t}(\mathcal{Q}_{n_b}) ,
%\end{equation}
%where $\bar{\varUpsilon}_{t}(\mathcal{Q}_{n_b})$ indicates the number of total backlogged and newly arrived requests before the end of slot $t$ and .

%\subsubsection{Computation Offloading at End Devices}
%
%\subsubsection{Dispatch of Requests in Edge Cluster}
%
%\subsubsection{Orchestration of Services at Cloud Cluster}
\vspace{-0.3em}
\begin{algorithm}
	\caption{\small Training and Scheduling Process of \textit{KaiS}}
	\label{algorithm:global}
	\small Initialize the system environment and neural networks.

%	\For{\rm{slot} $t$ = $\left \langle0, 1, ...\right \rangle$}{
	\For{\rm{slot} $t$ = $1, 2, ...$}{

		\If {\rm{frame} $\tau$ \rm{begins}}{
			
			Get reward $\tilde{u}_{\tau-1}$ and store $[\tilde{\boldsymbol{s}}_{\tau-1}, \tilde{\boldsymbol{a}}_{\tau-1}, \tilde{u}_{\tau-1}] $;
			
			Use GNNs to embed system states as Eq. (\ref{equ:gnn embedding});
			
			%Select action $\tilde{\boldsymbol{a}}_{\tau}$ using $\pi_{\theta^{*}}(\cdot)$.
			Select $H$ high-value edge nodes $(\tilde{a}_{\tau}^{\bullet})$ and
			\linebreak compute their service scaling actions $(\tilde{\boldsymbol{a}}_{\tau}^{\star})$  \hspace*{3.0em}%
			\hbox{\raise 5pt \rlap{\smash{$\left.\begin{array}{@{  }c@{}}\\{}\\{}\\{}\\{}\\{}\\{}\\{}\end{array}\color{orchestrator2}\right\}$}}}
			\hbox{\raise 9pt \rlap{\smash{$\color{orchestrator2}\begin{tabular}{l}\rotatebox{270}{\footnotesize \textbf{\textit{Orchestrate (GPG)}}}\end{tabular}$}}} 
			\linebreak using policy networks $\theta_g$ and $\theta_q$, respectively;

			Execute orchestration action $\tilde{\boldsymbol{a}}_{\tau} = (\tilde{a}_{\tau}^{\bullet}, \tilde{\boldsymbol{a}}_{\tau}^{\star})$;
			
			Update GNNs and policy networks using Eq. (\ref{equ:GNNDRL policy gradient});
		}
		
		 \ForPar{\rm{each eAP agent} $b \in B$ }{
			
			\If{ $\mathcal{Q}_{b} == \varnothing$}{
			
				\textbf{Continue}
			}
			
			Update request queue $\mathcal{Q}_{b}$ and get reward $\hat{u}_{b, t-1}$;
			
			Store $[ \hat{\boldsymbol{s}}_{b, t-1}, \hat{a}_{b, t-1}, A\left(\hat{\boldsymbol{s}}_{b, t-1}, \hat{a}_{b, t-1}\right),\hat{u}_{b, t-1},$
			\linebreak $ \boldsymbol{F}_{b, t-1}] $ for $\theta_{p}$ (actor);
			
			Dequeue current request $r_{b, t}$; 
			
			Compute the resource context $\boldsymbol{F}_{b, t}$ using Eq. (\ref{equ:resource context});\hspace*{1.0em}%
			\hbox{\raise -5pt \rlap{\smash{$\left.\begin{array}{@{}c@{}}\\{}\\{}\\{}\\{}\\{}\\{}\\{}\\{}\\{}\end{array}\color{dispatchers2}\right\}$}}}
			\hbox{\raise -1pt \rlap{\smash{$\color{dispatchers2}\begin{tabular}{l}\rotatebox{270}{\textbf{\footnotesize \textit{Dispatch (cMMAC)}}}\end{tabular}$}}}
			
			Take dispatch action $\hat{a}_{b, t}$ for $r_{b,t}$ using Eq. (\ref{equ:probability of valid actions for agent});
			
		}
		
		Store $[ \hat{\boldsymbol{s}}_{t-1}, V^{*}\left(\hat{\boldsymbol{s}}_{t-1} ; \theta_{v}^{\prime}, \pi\right)] $ for $\theta_{v}$ (critic);
		
		Update neural networks $\theta_{p}$ (actor) and $\theta_{v}$ (critic) 
		\linebreak centrally using Eq. (\ref{equ:the gradient of policy of cMAAC}) and  (\ref{equ:loss function derived from Bellman equation}), respectively; 
		
		Synchronize $\theta_{p}$ periodically to distributed eAPs.

	} 
\end{algorithm}
\vspace{-0.6em}

\vspace{-0.4em}
\section{Algorithm Design}
\label{sec:Algorithm and System Design}
\vspace{-0em}

The overall training and scheduling process of \textit{KaiS} is given in Algorithm \ref{algorithm:global}. 
We explain the technical details of request dispatch and service orchestration in the following.
Detailed training settings are presented in Sec. \ref{subsec:Training Settings}.

\vspace{-0.2em}
\subsection{Tailored MADRL for Decentralized Request Dispatch}
\vspace{-0.1em}

Request dispatch is to let each eAP independently decide which edge node or the cloud should serve the arrived request.
The goal of dispatch is to maximize the long-term system throughput rate $\varPhi^{\prime}$ by (\textit{$\romannumeral1$}) balancing the workloads among edge nodes and (\textit{$\romannumeral2$}) further offloading some requests to the cloud.

\subsubsection{Markov Game Formulation}

To employ MADRL, we formulate that eAPs independently perform request dispatch as a Markov game $\mathcal{G}$ for eAP agents $\mathcal{B} = \{ 1, 2, \ldots, B \}$.
Formally, the game $\mathcal{G} = ( \mathcal{B}, \hat{\mathcal{S}}, \hat{\mathcal{A}}, \hat{\mathcal{P}}, \hat{\mathcal{U}} )$ is defined as follows.
\begin{itemize}[leftmargin=*]
\item \textbf{\textit{State}}. 
$\hat{\mathcal{S}}$ is the state space.
At each slot $t$, we periodically construct a local state $\hat{\boldsymbol{s}}_{b, t}$ for each eAP agent $b$, which consists of (\textit{$\romannumeral1$}) the service type and delay requirement of the current dispatching request $r_{b, t}$, (\textit{$\romannumeral2$}) the queue information $\mathcal{Q}_{b, t}^{\prime}$ of requests awaiting dispatch at eAP $b$, (\textit{$\romannumeral3$}) the queue information, $\{ \mathcal{Q}_{n_b, t}^{\prime}:n_b \in \mathcal{N}_b \}$, of unprocessed requests at edge nodes $\mathcal{N}_b$, (\textit{$\romannumeral4$}) the remaining CPU, memory and storage resources of $\mathcal{N}_b$, (\textit{$\romannumeral5$}) the number of $\mathcal{N}_b$, i.e., $|\mathcal{N}_b| = N_{b}$, and (\textit{$\romannumeral6$}) the measured network latency between the eAP and the cloud.
Meanwhile, for centralized critic training, we maintain a global state $\hat{\boldsymbol{s}}_{t} \in \hat{\mathcal{S}}$, which includes (\textit{$\romannumeral1$}) the above information for all eAPs $\mathcal{B}$ and edge nodes $\mathcal{N}$, instead of only eAP $b$ and $\mathcal{N}_b$, and (\textit{$\romannumeral2$}) the queue information $\mathcal{Q}_{C, t}^{\prime}$ of unprocessed requests at the cloud cluster $C$.
\begin{figure}[t]%[!htp]
    \centering
%  \vspace{-0.8em}
    \includegraphics[width=8.95 cm]{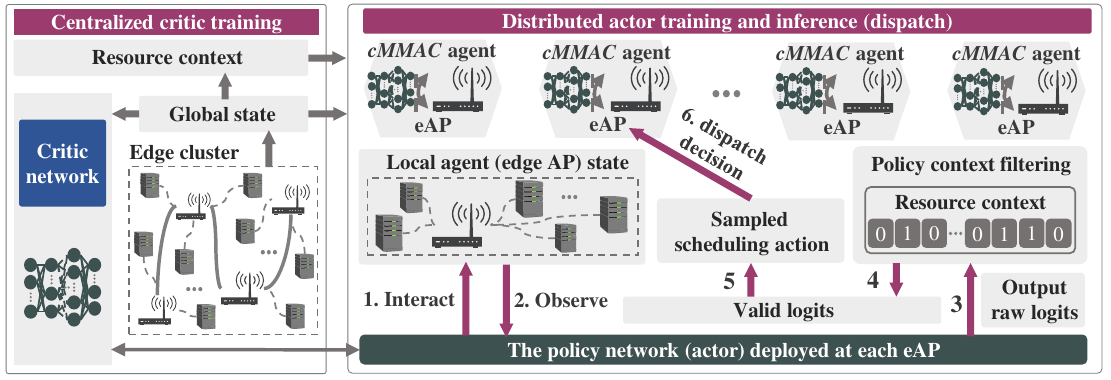}
\setlength{\abovecaptionskip}{-0.5cm} 
    \caption{Coordinated multi-agent actor-critic for decentralized request dispatch.}
  \vspace{-1.5em}
    \label{fig:cMMAC}
\end{figure}
\item \textbf{\textit{Action space}}. The joint action space of $\mathcal{G}$ is $\hat{\mathcal{A}}=\hat{\mathcal{A}}_{1} \times \ldots \times \hat{\mathcal{A}}_{b} \times \ldots \times \hat{\mathcal{A}}_{B}$, where the individual action space $\hat{\mathcal{A}}_{b}$ of eAP agent $b$ specifies where the current request can be dispatched to. For an edge cluster, we consider all available edge nodes as a resource pool, namely the cooperation between eAPs in enabled. In this case, $\hat{\mathcal{A}}_{b}$ includes $N+1$ discrete actions denoted by ${\{i\}}_{0}^{N}$, where $\hat{a}_{b, t} = 0$ and $\hat{a}_{b, t} \in \mathcal{N}$ specify dispatching to the cloud or edge nodes, respectively. At each slot $t$, we use $ \hat{\boldsymbol{a}}_{t} = ( \hat{a}_{b, t}: b \in \mathcal{B} ) $ to represent the joint dispatch actions of all requests required to be scheduled at all eAPs. Note that multiple requests may be queued in eAPs ($\mathcal{Q}_{b}, b \in \mathcal{B}$), we only allow each eAP agent to dispatch one request at a slot $t$. Meanwhile, for \textit{KaiS}, we set the time slot to a moderate value (refer to Sec. \ref{sec:System Design and Implementation}) to ensure its scheduling timeliness to serve request arrivals.
\item \textbf{\textit{Reward function}}. All agents in the same edge cluster share a reward function $\hat{U} = \hat{\mathcal{S}} \times \hat{\mathcal{A}} \rightarrow \mathbb{R} $, i.e., $\hat{U}_b = \hat{U}$ holds for all $b \in \mathcal{B}$. Each agent wants to maximize its own expected discounted return $\mathbb{E}\left[\sum\nolimits_{i=0}^{\infty} \gamma^{i} \hat{u}_{b, t+i}\right]$, where $\hat{u}_{b, t}$ is the immediate reward for the $b$-th agent associated with the action $\hat{a}_{b, t}$ and $\gamma \in (0,1]$ is a discount factor. The immediate reward is defined as $\hat{u}_{b, t} = {\text{e}}^{-\lambda - \varepsilon \nu }$. Specifically, (\textit{$\romannumeral1$}) $\lambda \in [0, 1]$ is the ratio of requests that violate delay requirements during $[t, t+1]$, (\textit{$\romannumeral2$}) $\nu = 1 / (1+\mathrm{e}^{-\xi}) \in [0.5, 1]$, where $\xi \in \mathbb{R}_{\ge 0}$ is the standard deviation of the CPU and memory usage of all edge nodes, and (\textit{$\romannumeral3$}) $\varepsilon$ is the weight to control the degree of load balancing among edge nodes. The introduction of $\nu$ is to stabilize the system, preventing too much load are imposed on some edge nodes. When $\nu$ is closer to $0.5$, i.e., $\xi \to 0$, the loads of edge nodes are more balanced, thus leading to more scheduling rooms for dispatch. Such a reward is to \textit{improve the long-term throughput while ensuring the load balancing at the edge}. 
\item \textbf{\textit{State transition probability}}. We use $ p \left( \hat{\boldsymbol{s}}_{t+1} \mid \hat{\boldsymbol{s}}_{t}, \hat{\boldsymbol{a}}_{t} \right): \hat{\mathcal{S}} \times \hat{\mathcal{A}} \times \hat{\mathcal{S}} \rightarrow [0, 1] $ to indicate the transition probability from state $\hat{\boldsymbol{s}}_{t}$ to $\hat{\boldsymbol{s}}_{t+1}$ given a joint dispatch action $\hat{\boldsymbol{a}}_{t}$. The action is deterministic in $\mathcal{G}$, i.e., if $\hat{a}_{b, t} = 2$, the agent $b$ will dispatch the current request to edge node $2$ at slot $t+1$. 
\end{itemize}

\subsubsection{Coordinated Multi-Agent Actor-Critic}
\label{subsubsec:Coordinated Multi-agent Actor-Critic}

The challenges of training these dispatch agents are:
(\textit{$\romannumeral1$}) The environment of each agent is non-stationary because other agents are learning and affecting the environment simultaneously. 
Specifically, each agent usually learns its own policy that is changing over time \cite{MARLSurvey}, increasing the difficulty of coordinating them;
(\textit{$\romannumeral2$}) The action space of each agent changes dynamically since its feasible dispatch options vary with the available system resources, making vanilla DRL algorithms unable to handle. 
For instance, if the memory of the edge node $n$ is run out at slot $t$, the dispatch action ${\hat{a}}_{b, t+1}$ should not include that dispatching the request to edge node $n$.

Therefore, we design \textit{coordinated Multi-Agent Actor-Critic (cMMAC)}, as illustrated in Fig.~\ref{fig:cMMAC}: 
(\textit{$\romannumeral1$}) Adopt a centralized critic and distributed actors to coordinate learning, i.e., all agents share a centralized state-value function when training critic, while during distributed actor training and inference each actor only observes the local state. 
(\textit{$\romannumeral2$}) By policy context filtering, we can adjust their policies to tolerate dynamic action space and establish explicit coordination among agents to facilitate successful training.
The details are illustrated as follows.

\begin{itemize}[leftmargin=*]
\item \textbf{\textit{Centralized state-value function (Critic)}}. The state-value function shared by eAP agents can be obtained by minimizing the loss function derived from Bellman equation \cite{sutton2018reinforcement}:
\vspace{-0.7em}
\begin{equation}
\label{equ:loss function derived from Bellman equation}
L\left(\theta_{v}\right)=\left(V_{\theta_{v}}\left(\hat{\boldsymbol{s}}_{b, t}\right)-V^{\ast}\left(\hat{\boldsymbol{s}}_{t+1} ; \theta_{v}^{\prime}, \pi\right)\right)^{2},
\end{equation}
\begin{multline}
\label{equ:centralized state-value function}
V^{\ast}\left(\hat{\boldsymbol{s}}_{t+1}; \theta_{v}^{\prime}, \pi\right)= \\ \sum\nolimits_{\hat{a}_{b, t}} \pi\left(\hat{a}_{b,t} \mid \hat{\boldsymbol{s}}_{b, t}\right)\left(\hat{u}_{b, t+1}+\gamma V_{\theta_{v}^{\prime}}\left(\hat{\boldsymbol{s}}_{b, t+1}\right)\right),
\end{multline}
%\begin{itemize}[leftmargin=*]
%\item \textbf{\textit{Centralized state-value function (Critic)}}. The centralized state-value function shared by eAP agents can be obtained by minimizing the following loss function derived from Bellman equation \cite{sutton2018reinforcement}:
%\vspace{0.5em}
%\begin{equation}
%\label{equ:loss function derived from Bellman equation}
%L\left(\theta_{v}\right)=\left(V_{\theta_{v}}\left(\hat{\boldsymbol{s}}_{t}\right)-V^{\ast}\left(\hat{\boldsymbol{s}}_{t+1} ; \theta_{v}^{\prime}, \pi\right)\right)^{2},
%\end{equation}
%\begin{multline}
%\label{equ:centralized state-value function}
%V^{\ast}\left(\hat{\boldsymbol{s}}_{t+1}; \theta_{v}^{\prime}, \pi\right)= \\ \sum\nolimits_{\hat{a}_{b, t}} \pi\left(\hat{a}_{b,t} \mid \hat{\boldsymbol{s}}_{t}\right)\left(\hat{u}_{b, t+1}+\gamma V_{\theta_{v}^{\prime}}\left(\hat{\boldsymbol{s}}_{t+1}\right)\right),
%\end{multline}
where $\theta_{v}$ and $\theta_{v}^{\prime}$ denote the parameters of the value network and the target value network, respectively.
In total, for $B$ eAP agents, there are $B$ unique state-values $\{ V(\hat{\boldsymbol{s}}_{b, t}): b \in \mathcal{B} \}$ at each slot. 
Each state-value output $V(\hat{\boldsymbol{s}}_{b, t})$ is the expected return received by agent $b$ at slot $t$.
To stabilize the learning of the state-value function, we fix a target value network $V^{\ast}$ parameterized by $\theta_{v}^{\prime}$ and update it at the end of each training episode.
\item \textbf{\textit{Policy context filtering (Actors)}}. Policy context filtering is mainly reflected in the resource context when scheduling request dispatch. In the operating edge-cloud system, the available resources of edge nodes fluctuate along with the scheduling events. To avoid, as much as possible, the situation that an eAP agent dispatches a request to an edge node with insufficient resources, before dispatch, we compute a resource context $\boldsymbol{F}_{b, t} \in \{0, 1\}^{N+1}$ for each eAP agent, which is a binary vector that filters out invalid dispatch actions. The value of the element of $\boldsymbol{F}_{b, t}$ is defined as:
\begin{equation}
\label{equ:resource context}
\left[\boldsymbol{F}_{b, t}\right]_{j}=\left\{\begin{array}{ll}
1, & \text{ if edge node } j \text{ is available,} \\
1, & \text{ if } j = 0, \\
0, & \text { otherwise, }
\end{array}\right.
\end{equation}
where (\textit{$\romannumeral1$}) $[\boldsymbol{F}_{b, t}]_{j}$ ($j = 1, \ldots, N$) represents the validity of dispatching the current request to $j$-th edge node and (\textit{$\romannumeral2$}) $[\boldsymbol{F}_{b, t}]_{0}$ specifies that the cloud cluster ($j=0$) is always a valid action of request dispatch, namely $\left[\boldsymbol{F}_{b, t}\right]_{0} \equiv 1$.
The coordination of agents is also achieved by masking available action space based on the resource context $\boldsymbol{F}_{b, t}$.
To proceed, we first use $\boldsymbol{p}(\hat{\boldsymbol{s}}_{b,t}) \in \mathbb{R}^{N+1}$ to denote the original output logits from the actor policy network for the $b$-th agent conditioned on state $\hat{\boldsymbol{s}}_{b,t}$. Then, we let $\bar{\boldsymbol{p}}(\hat{\boldsymbol{s}}_{b, t})=\boldsymbol{p}(\hat{\boldsymbol{s}}_{b, t}) * \boldsymbol{F}_{b, t}$, where the operation $*$ is element-wise multiplication, to denote the valid logits considering the resource context for agent $b$. Note that the output logits $\boldsymbol{p}(\hat{\boldsymbol{s}}_{b,t}) \in \mathbb{R}^{N+1}_{>0}$ are restricted to be positive to achieve effective masking. Based on the above denotations, the probability of valid dispatch actions for agent $b$ can be given by:
\begin{equation}
\label{equ:probability of valid actions for agent}
\pi_{\theta_{p}}\left(\hat{a}_{b,t}=j \mid \hat{\boldsymbol{s}}_{b,t}\right)=\left[\bar{\boldsymbol{p}}\left(\hat{\boldsymbol{s}}_{b,t}\right)\right]_{j}=\frac{[\bar{\boldsymbol{p}}(\hat{\boldsymbol{s}}_{b,t})]_{j}}{\|\bar{\boldsymbol{p}}(\hat{\boldsymbol{s}}_{b, t})\|_{1}}, 
\end{equation}
where $\theta_{p}$ is the parameters of actor policy network. At last, for \textit{cMMAC}, the policy gradient $\nabla_{\theta_{p}} J(\theta_{p})$ can be derived and the advantage $A(\hat{\boldsymbol{s}}_{b, t}, \hat{a}_{b, t})$ can be computed as follows:
\begin{equation}
\label{equ:the gradient of policy of cMAAC}
\nabla_{\theta_{p}} J\left(\theta_{p}\right)=\nabla_{\theta_{p}} \log \pi_{\theta_{p}}\left(\hat{a}_{b, t} \mid \hat{\boldsymbol{s}}_{b, t}\right) A\left(\hat{\boldsymbol{s}}_{b, t}, \hat{a}_{b, t}\right),
\end{equation}
\begin{equation}
\label{equ:the advantage of cMAAC}
A\left(\hat{\boldsymbol{s}}_{b, t}, \hat{a}_{b, t}\right)=\hat{u}_{b, t+1}+\gamma V_{\theta_{v}^{\prime}}\left(\hat{\boldsymbol{s}}_{b, t+1}\right)-V_{\theta_{v}}\left(\hat{\boldsymbol{s}}_{b, t}\right).
\end{equation}

\end{itemize}

\vspace{-0.3em}
\subsection{GNN-based Learning for Service Orchestration}
\label{subsec:GNN-based DRL for Service Orchestration}
\vspace{-0.2em}

We propose a \textit{GNN-based Policy Gradient (GPG)} algorithm and describe how 
(\textit{$\romannumeral1$}) the system state information is processed flexibly;
(\textit{$\romannumeral2$}) the high-dimensional service orchestration is decomposed as stepwise scheduling actions: selecting high-value edge nodes and then performing service scaling on them.
%(\textit{$\romannumeral3$}) the training is achieved under the continuous stochastic requests.

\subsubsection{GNN-based System State Encoding}
\label{subsubsec:GNN-based System State Encoding}

As shown in Fig.~\ref{fig:GNNEncode}, \textit{KaiS} must convert system states into feature vectors on each observation and then pass them to policy networks.
A common choice is directly stacking system states into flat vectors.
However, the edge-cloud system is practically a graph consisting of connected eAPs, edge nodes, and the cloud cluster.
Simply stacking states has two defects: 
(\textit{$\romannumeral1$}) processing a high-dimensional feature vector requires sophisticated policy networks, which increases training difficulty; 
(\textit{$\romannumeral2$}) it cannot efficiently model the graph structure information for the system, making \textit{KaiS} hard to generalize to various system scales and structures.
Therefore, we use GNNs to encode system states into a set of embeddings layer by layer as follows.
%\vspace{-0.4em}
\begin{itemize}[leftmargin=*]
\item \textbf{\textit{Embedding of edge nodes}}.
For edge nodes associated with eAP $b$, each of them, $n_b \in \mathcal{N}_b$, carries the following attributes at each frame $\tau$, denoted by a vector $\tilde{\boldsymbol{s}}_{n_b, \tau}$: (\textit{$\romannumeral1$}) the available resources of CPU, memory, storage, etc., (\textit{$\romannumeral2$}) the periodically measured network latency with eAP $b$ and the cloud, (\textit{$\romannumeral3$}) the queue information of the backlogged requests at itself, i.e., $\mathcal{Q}_{n_b}^{\prime}$, and (\textit{$\romannumeral4$}) the indexes of deployed services and the number of replicates of each deployed service. 
Given $\tilde{\boldsymbol{s}}_{n_b, \tau}$, \textit{KaiS} performs embedding for each edge node as $ (\mathcal{N}_b, \tilde{\boldsymbol{s}}_{n_b, \tau}) \rightarrow \boldsymbol{x}_{n_b, \tau}$. 
To perform embedding, for an edge node $n_b \in \mathcal{N}_b$, we build a virtual graph by treating other edge nodes $\mathcal{N}_b \setminus n_b$ as its neighbor nodes.
Then, as depicted in Fig.~\ref{fig:GNNEncode}, we traverse the edge nodes in $\mathcal{N}_b$ and compute their embedding results one by one. 
Once an edge node has accomplished embedding, it provides only the embedding results $\boldsymbol{x}_{n_b, \tau}$ for the subsequent embedding processes of the remaining edge nodes.
For edge node $n_b \in \mathcal{N}_b$, its embedding results $\boldsymbol{x}_{n_b, \tau}$ can be computed by propagating information from its neighbor nodes $\zeta(n_b) = \{ \mathcal{N}_b \setminus n_b \}$ to itself in a message passing step.
In message passing, edge node $n_b$ aggregates messages from all of its neighbor nodes and computes its embeddings as:
\begin{equation}
\label{equ:gnn embedding}
\boldsymbol{x}_{n_b, \tau}=h_{1}\Big[\sum\nolimits_{n_b^{\prime} \in \zeta(n_b)} f_{1}(\boldsymbol{x}_{n_b^{\prime}, \tau})\Big]+\tilde{\boldsymbol{s}}_{n_b, \tau},
\end{equation}
where $h_1(\cdot)$ and $f_1(\cdot)$ are both non-linear transformations implemented by Neural Networks (NNs), combined to express a wide variety of aggregation functions.
Throughout the embedding, we reuse the same NNs $h_1(\cdot)$ and $f_1(\cdot)$.
\begin{figure}[t]%[!htp]
    \centering
%  \vspace{-0.8em}
    \includegraphics[width=8.85 cm]{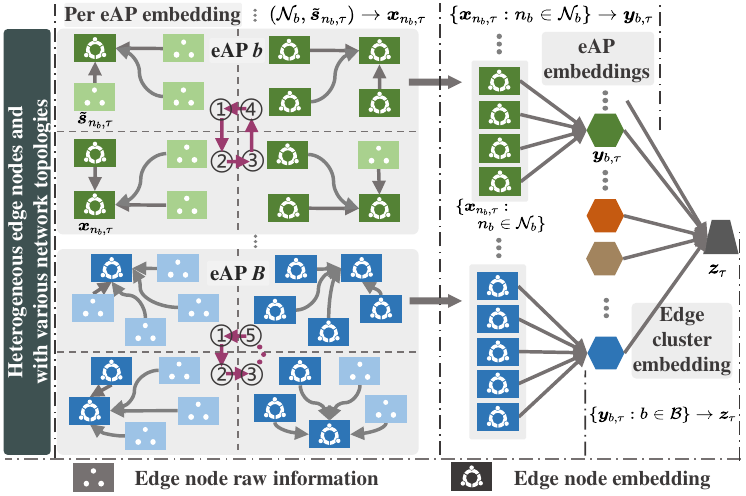}
\setlength{\abovecaptionskip}{-0.5cm} 
    \caption{GNN-based system state encoding for the edge cluster.}
  \vspace{-1em}
    \label{fig:GNNEncode}
\end{figure}
\item \textbf{\textit{Embedding of eAPs and the edge cluster}}. Similarly, we leverage GNNs to compute an eAP embedding for each eAP $b$, $ \{ \boldsymbol{x}_{n_b, \tau}: n_b \in \mathcal{N}_b \} \rightarrow \boldsymbol{y}_{b, \tau}$, and further an edge cluster embedding for all eAPs, $ \{\boldsymbol{y}_{b, \tau}: b \in \mathcal{B} \} \rightarrow \boldsymbol{z}_{\tau} $. 
To compute the embedding for eAP $b$ as in (\ref{equ:gnn embedding}), we add an eAP summary node to $\mathcal{N}_b$ and treat all edge nodes in $\mathcal{N}_b$ as its neighbor nodes.
These eAP summary nodes are also used to store their respective eAP embeddings.
Then, the eAP embedding for each eAP can be obtained by aggregating messages from all neighboring nodes and computed as (\ref{equ:gnn embedding}).
In turn, these eAP summary nodes are regarded as the neighbor nodes of an edge cluster summary node, such that (\ref{equ:gnn embedding}) can be used to compute the global embedding as well. 
Though the embeddings $\boldsymbol{y}_{b, \tau}$ and $\boldsymbol{z}_{\tau}$ are both computed by (\ref{equ:gnn embedding}), different sets of NNs, i.e., (\textit{$\romannumeral1$}) $h_2(\cdot)$, $f_2(\cdot)$ for $\boldsymbol{y}_{b, \tau}$ and (\textit{$\romannumeral2$}) $h_3(\cdot)$, $f_3(\cdot)$ for $\boldsymbol{z}_{\tau}$, are used for non-linear transformations.
 \end{itemize} 

%\vspace{-0.1em}
\subsubsection{Stepwise Scheduling for Service Orchestration}

The key challenge in encoding service orchestration actions is to deal with the learning and computational complexity of high-dimensional action spaces. 
A direct solution is to maintain a huge policy network and orchestrate all services $\mathcal{W}$ for all edge nodes $\mathcal{N}$ at once based on the embedding results in Sec. \ref{subsubsec:GNN-based System State Encoding}. 
However, in this manner, \textit{KaiS} must choose actions from a large set of combinations $( \tilde{d}_{w, n} \in \mathbb{N}: w \in \mathcal{W}, n \in \mathcal{N})$, thereby increasing sample complexity and slowing down training \cite{silver2016mastering}. 
Besides, too frequent large-scale service orchestration will bring huge system overhead and harm system stability. 

Therefore, we consider stepwise scheduling, which in each frame first selects $H$ high-value edge nodes ($H=2$ in experiments), and then scales services for each of them in a customized action space of a much smaller size $2M+1$. 
%The action space size of this approach, $H(2M+1)$, is much smaller, and involving only a few edge nodes at a frame will not affect the stability of the system.
%\textit{KaiS} chooses to orchestrate services at some edge nodes at each frame $\tau$. 
Specifically, \textit{KaiS} passes the embedding vectors from Sec. \ref{subsubsec:GNN-based System State Encoding} as inputs to the policy networks, which output a joint orchestration action $\tilde{\boldsymbol{a}}_{\tau} = (\tilde{a}_{\tau}^{\bullet}, \tilde{\boldsymbol{a}}_{\tau}^{\star})$, including (\textit{$\romannumeral1$}) the action of selecting high-value edge nodes $\tilde{a}_{\tau}^{\bullet}$ and (\textit{$\romannumeral2$}) the joint service scaling action $\tilde{\boldsymbol{a}}_{\tau}^{\star}$ corresponding to high-value edge nodes.
\begin{figure}[t]%[!htp]
    \centering
%  \vspace{-0.6em}
    \includegraphics[width=8.85 cm]{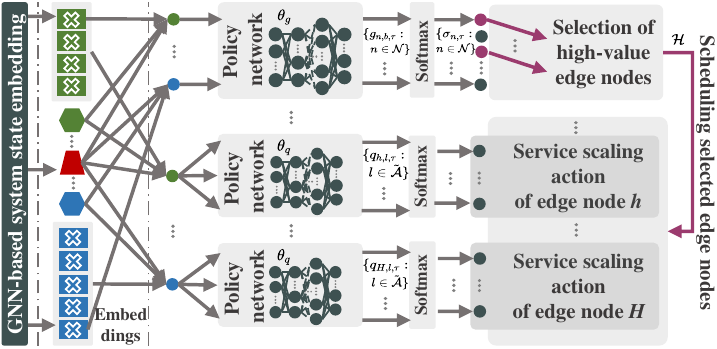}
\setlength{\abovecaptionskip}{-0.5cm} 
    \caption{GNN-based learning, i.e., \textit{GPG}, for service orchestration.}
  \vspace{-1.3em}
    \label{fig:GPG}
\end{figure}
\begin{itemize}[leftmargin=*]
\item \textbf{\textit{Selection of high-value edge nodes}}. At each frame, \textit{KaiS} first uses a policy network to select $H (\leq N)$ high-value edge nodes, denoted by action $\tilde{a}_{\tau}^{\bullet}$. As illustrated in Fig.~\ref{fig:GPG}, for edge node $n$ associated with eAP $b$, it computes a value $g_{n, b, \tau} = g(\boldsymbol{x}_{n, \tau}, \boldsymbol{y}_{b, \tau}, \boldsymbol{z}_{\tau})$, where $g(\cdot)$ is a non-linear value-evaluation function implemented by a NN $\theta_g$. The introduction of function $g(\cdot)$ is to map the embedding vectors to a scalar value. The value $g_{n, b, \tau}$ specifies the priority of \textit{KaiS} performing service scaling at edge node $n$. A softmax operation is used to compute the probability $\sigma_{n, \tau}$ of selecting edge node $n$ based on the values $\{ g_{n, b, \tau}: n \in \mathcal{N} \}$:
\begin{equation}
\label{equ:softmax}
\sigma_{n, \tau} = \mathrm{e}^{g_{n, b, \tau}} / \sum\nolimits_{ b^{\prime} \in \mathcal{B} } \sum\nolimits_{n_{b^{\prime}} \in \mathcal{N}_{b^{\prime}}} \mathrm{e}^{g_{n_{b^{\prime}}, b^{\prime}, \tau}}.
\end{equation}
According to the probabilities $\{ \sigma_{n, \tau}: n \in \mathcal{N} \}$ for all edge nodes, \textit{KaiS} selects $H$ edge nodes with high probabilities as high-value edge nodes $\mathcal{H}$ to perform service scaling.
\item \textbf{\textit{Service scaling for high-value edge nodes}}. 
For a selected high-value edge node $h \in \mathcal{H} \subset \mathcal{N}$, \textit{KaiS} uses an action-evaluation function $q(\cdot)$, implemented by a NN $\theta_q$, to compute a value $q_{h, l, \tau} = q(\boldsymbol{x}_{h, \tau}, \boldsymbol{y}_{b, \tau}, \boldsymbol{z}_{\tau}, l)$ for edge node $h$ performing service scaling $\tilde{a}_{h, \tau} = l$ at frame $\tau$.
The action space of $l$ is defined as $\tilde{\mathcal{A}} \triangleq (-W, \ldots, w, \ldots, W)$ with size $2W+1$, i.e., $l \in \tilde{\mathcal{A}}$.
The meaning of $l$ is as follows: (\textit{$\romannumeral1$}) $l = 0$ indicates that all services remain unchanged, (\textit{$\romannumeral2$}) $l = -w$ means deleting a replicate of service $w$, and (\textit{$\romannumeral3$}) $l = w$ specifies adding a replicate of service $w$.
Particularly, for an invalid service scaling action due to resource limitations of an edge node, \textit{KaiS} always transforms it to $l = 0$. 
Similarly, we apply a softmax operation on $\{ q_{h, l, \tau}: l \in \tilde{\mathcal{A}} \}$ to compute the probabilities of scaling actions, and choose to perform the action with the highest probability.
For all high-value edge nodes $\mathcal{H}$, \textit{KaiS} will generate a joint service scaling action $\tilde{\boldsymbol{a}}_{\tau}^{\star} = ( \tilde{a}_{h, \tau}: h \in \mathcal{H} )$ at each frame.
\end{itemize}

While \textit{KaiS} decouples request dispatch and service orchestration, this does not affect our objective of improving $\varPhi^{\prime}$.
In fact, as the dispatch policy is contained in a regularly updated policy network that provides an appropriate load-balanced edge cluster for orchestration, we also implicitly optimize the dispatch when optimizing the orchestration.

To guide \textit{GPG}, \textit{KaiS} generates a reward $\tilde{u}_{\tau} = \mathrm{e}^{ - \sum\nolimits_{n \in \mathcal{N}} \mid \mathcal{Q}_{n, \tau} \mid } $ after each service orchestration at frame $\tau$, where $\mid \mathcal{Q}_{n, \tau} \mid$ is the queue length of unprocessed requests at edge node $n$.
By such a design, \textit{GPG} will gradually lead \textit{KaiS} to reduce backlogged requests as much as possible, thereby improving the throughput rate, which we will show in experiments.
\textit{KaiS} adopts a policy gradient algorithm for training NNs $\{ f_{i}(\cdot), h_{i}(\cdot) \}_{i=1,2,3}$, $\theta_g$ and $\theta_q$ used in \textit{GPG}.
For clarity, we denote all parameters of these NNs jointly as $\theta^{*}$, all GNN-encoded system states as $\tilde{\boldsymbol{s}}_{\tau}$, the joint service orchestration action as $\tilde{\boldsymbol{a}}_{\tau}$, and the scheduling policy as $\pi_{\theta^{*}}\left( \tilde{\boldsymbol{s}}_{\tau}, \tilde{\boldsymbol{a}}_{\tau} \right)$, i.e., the probability of taking action $\tilde{\boldsymbol{a}}_{\tau}$ when observing state $\tilde{\boldsymbol{s}}_{\tau}$.
At each frame, \textit{KaiS} collects the observation $(\tilde{\boldsymbol{s}}_{\tau}, \tilde{\boldsymbol{a}}_{\tau}, \tilde{u}_{\tau})$ and updates the parameters $\theta^{*}$ using policy gradient:
\begin{equation}
\label{equ:GNNDRL policy gradient}
\small
\theta^{*} \leftarrow \theta^{*} +\alpha \sum_{\tau=1}^{T} \nabla_{\theta^{*}} \log \pi_{\theta^{*}}\left(\tilde{\boldsymbol{s}}_{\tau}, \tilde{\boldsymbol{a}}_{\tau}\right)\left(\sum_{\tau^{\prime}=\tau}^{T} \tilde{u}_{\tau^{\prime}}-\mu_{\tau}\right),
\end{equation}
where $T$ is the length of a \textit{GPG} training episode, $\alpha$ is the learning rate, and $\mu_{\tau}$ is a baseline used to reduce the variance of the policy gradient. 
A method for computing the baseline is setting $\mu_{\tau}$ to the cumulative reward from frame $\tau$ onwards, averaged over all training episodes \cite{greensmith2004variance}.

\vspace{-0.2em}
\section{Implementation Design}
\label{sec:System Design and Implementation}

All services are hosted in the system as \textit{Docker} containers.
\textit{KaiS} is implemented based on \textit{k8s} and \textit{k3s} (a lightweight \textit{k8s} for edge) \cite{k3s} in Ubuntu 16.04 using Python 3.6.
\vspace{-0.5em}
\subsection{Edge-cloud System Setup}
\label{subsec:End-edge-cloud Computing System Setup}
\vspace{-0.2em}

\begin{itemize}[leftmargin=*]
\item \textbf{\textit{Requests}}. 
Real-world workload traces from Alibaba \cite{AlibabaDatasets} are modified and used to generate service requests. 
We classify the workload requests in that trace into $30$ services.
Specifically, the ``task\_type'' in that trace is considered the service type and the delay requirements of each request are acquired by properly scaling ``start\_time'' minus ``end\_time''.
Instead of employing real end devices, we implement a \textit{request generator} to generate service requests and then forward them to \textit{k3s master nodes} (eAPs) at random. 
\item \textbf{\textit{Edge cluster and nodes}}. 
By default, we set up $5$ \textit{k3s} clusters in different geographic regions of the Google Cloud Platform (GCP) to emulate geographic distribution, each cluster consisting of a \textit{k3s master node} and $8$ \textit{k3s edge nodes}.
\textit{K3s master nodes} and \textit{k3s edge nodes} use GCP Virtual Machine (VM) configurations ``2 vCPU, 4 GB memory, and 0.3 TB disk'' and ``1-2 vCPU, 2-4 GB memory, and 0.3 TB disk'', respectively. 
Besides, we use more powerful \textit{k3s master nodes} to accelerate offline training.
%during offline training, we use more powerful \textit{k3s edge nodes} to accelerate training.
\item \textbf{\textit{Cloud cluster}}. 
We build a homogeneous 15-VM cluster as the cloud cluster, where each VM is with ``4 vCPU, 16 GB memory, and 1 TB disk''. 
A \textit{k8s master node} is deployed at one VM to manage others. 
We handcraft $30$ services with various CPU and memory consumption and store their Docker images in the cloud. 
Moreover, we intentionally control the network bandwidth and delay between the cloud and the edge, with \textit{Linux TC}, to simulate practical scenarios.  
\end{itemize}

\vspace{-0.6em}
\subsection{Main Components of KaiS}
\label{subsec:Main Components of KaiS}
\vspace{-0.25em}

\begin{figure}[t]%[!htp]
    \centering
%  \vspace{-0.4em}
    \includegraphics[width=8.85 cm]{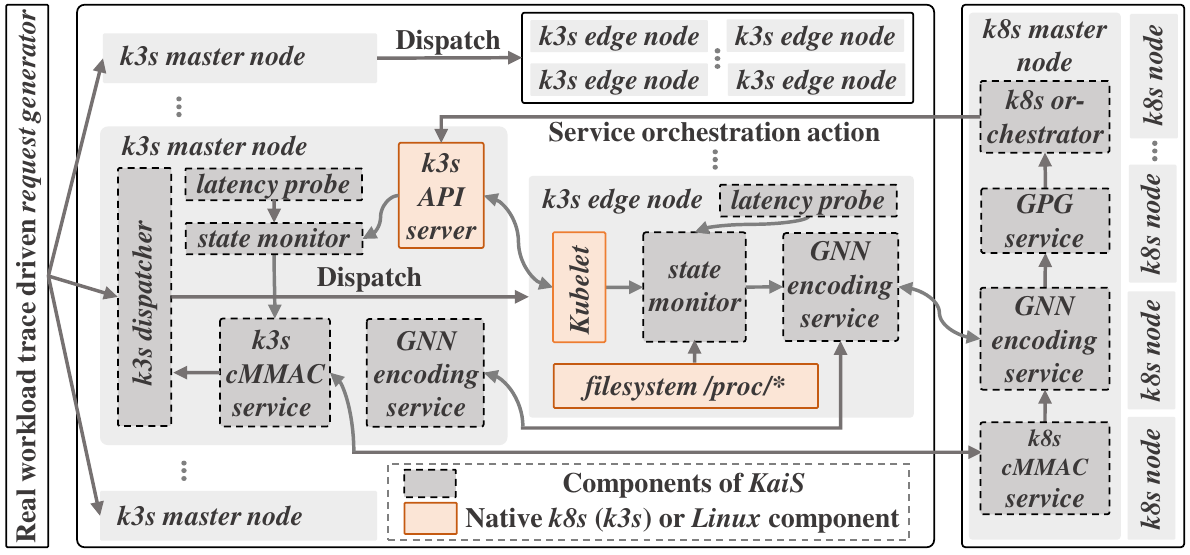}
\setlength{\abovecaptionskip}{-0.5cm} 
    \caption{Implementation and prototype design of \textit{KaiS}.}
  \vspace{-1.3em}
    \label{fig:ImplementationDesign}
\end{figure}

We decouple \textit{KaiS} into two main parts as shown in Fig.~\ref{fig:ImplementationDesign}. 
\begin{itemize}[leftmargin=*]
\item \textbf{\textit{Decentralized request dispatchers}}. 
\textit{KaiS} maintains a \textit{k3s dispatcher} at each \textit{k3s master node} to periodically observe and collect the current system states by a \textit{state monitor} in the following manner. 
Each \textit{k3s edge node} (\textit{$\romannumeral1$}) runs a \textit{Kubelet} process and (\textit{$\romannumeral2$}) reads the virtual \textit{filesystem /proc/*} in \textit{Linux} to collect the states about \textit{Docker} services and physical nodes.
Concerning network status, each \textit{k3s edge node} and \textit{k3s master node} host a \textit{latency probe} to measure network latency.
\textit{State monitors} at \textit{k3s edge nodes} will periodically push the above collected system states to the \textit{state monitor} at the \textit{k3s master node} for fusion.
To implement \textit{cMMAC}, we deploy a \textit{cMMAC} agent at each \textit{k3s master node} as \textit{k3s cMMAC service} while maintaining a \textit{k8s cMMAC service} at the \textit{k8s master node} to support training. 
At each scheduling slot $0.25$~s, empirically determined from experiment results as shown in Fig.~\ref{fig:HyperParameters}, the \textit{k3s cMMAC service} at a \textit{k3s master node} computes a dispatch action by observing local states from the \textit{state monitor}, and then notifies the \textit{k3s dispatcher} to execute this dispatch for the current request.
\item \textbf{\textit{Centralized service orchestrator}}. To implement \textit{GPG}, \textit{KaiS} holds a set of \textit{GNN encoding services} with different GNNs (Sec. \ref{subsubsec:GNN-based System State Encoding}) at \textit{k3s edge nodes}, \textit{k3s master nodes} and \textit{k8s master node}.
These \textit{GNN encoding services} are communicated with each other and used to compute the embeddings of edge nodes, eAPs (i.e., \textit{k3s master nodes}) and the edge cluster, respectively. 
Once \textit{KaiS} finishes the GNN-based encoding, the \textit{GNN encoding service} at \textit{k8s master node} will merge all embedding results.
The remaining parts of \textit{GPG}, i.e., the policy networks, are realized as a \textit{GPG service} and deployed at the \textit{k8s master node}.
Frame length is set as $100\times$ slots to ensure system stability.
At each frame, the \textit{GPG service} pulls all embeddings from the \textit{GNN encoding service} and computes the orchestration action.
Then, the \textit{GPG service} calls the \textit{k8s orchestrator} to communicate with specific \textit{k3s API servers} to accomplish service scaling via \textit{python-k8sclient}.
Unlike other scaling actions, only when a service is idle, the \textit{k3s API server} can delete it.
Otherwise, \textit{KaiS} will delay the scaling until the condition is met.
\end{itemize}

\vspace{-0.5em}
\subsection{Training Settings}
\label{subsec:Training Settings}

We implement Algorithm \ref{algorithm:global} based on TensorFlow 1.14.
The detailed settings are as follows.
\textbf{\textit{cMMAC}}: 
\textit{cMMAC} involves a critic network and an actor policy network. 
Both of them are trained using Adam optimizer with a fixed learning rate of $5\times10^{-4}$.
The critic $\theta_v$ is parameterized by a four-layer ReLU NN, where the node sizes of each layer are 256, 128, 64 and 32, respectively. 
The actor $\theta_p$ is implemented using a three-layer ReLU NN, with 256, 128, and 32 hidden units on each layer.
Note that the output layer of the actor uses ReLU+1 as an activation function to ensure that the elements in the original logits are positive. 
\textbf{\textit{GPG}}: 
\textit{GPG} uses (\textit{$\romannumeral1$}) six GNNs, i.e., $\{ f_{i}(\cdot), h_{i}(\cdot) \}_{i=1,2,3}$ and (\textit{$\romannumeral2$}) two policy networks including $\theta_g$ and $\theta_q$. 
Among them, $\{ f_{i}(\cdot), h_{i}(\cdot) \}_{i=1,2,3}$ are implemented with two-hidden-layer NNs with 64 and 32 hidden units on each layer. 
Besides, $\theta_g$ and $\theta_q$ are both three-hidden-layer NNs with node sizes of 128, 64 and 32 from the first layer to the last layer.
All NNs, related to \textit{GPG}, use Adam optimizer with a learning rate of $10^{-3}$ for parameter updates.

%\vspace{-0.2em}
\section{Performance Evaluation}
\label{sec:Performance Evaluation}
%We next evaluate our design and prototype implementation of \textit{KaiS} in the following aspects: 
%(\textit{$\romannumeral1$}) The learning ability of \textit{KaiS} adapting to request sequences with various patterns, while with moderate scheduling delay for request dispatch; 
%(\textit{$\romannumeral2$}) The impact of load-balancing degree in the edge cluster on the performance of \textit{KaiS}; 
%(\textit{$\romannumeral3$}) The role of GNN-based service orchestration, i.e., \textit{GPG} algorithm, to cope with pattern-fluctuating request sequences and different system scales;
%(\textit{$\romannumeral4$}) The superiority of \textit{KaiS} against other baseline scheduling methods in addressing stochastic arriving sequences.

In our evaluation, baseline scheduling methods include: 
(\textit{$\romannumeral1$}) \textbf{\textit{Greedy}} (for dispatch), which schedules each request to the edge node with the lowest resource utilization;
(\textit{$\romannumeral2$}) \textbf{\textit{Native}} (for orchestration), i.e., the default \textit{Horizontal Pod Autoscaler} \cite{k8sNativeScheduler} in \textit{k8s} based on the observation of specific system metrics; 
(\textit{$\romannumeral3$}) \textbf{\textit{GSP-SS}}\cite{Farhadi2019} (for both), assuming that the request arrival rate of each service is known in advance; 
(\textit{$\romannumeral4$}) \textbf{\textit{Firmament}}\cite{Gog2016} (for dispatch), designed to find the policy-optimal assignment of work (request) to cloud cluster resources.

We consider three main performance metrics:
%(\textit{$\romannumeral1$}) \textbf{\textit{Effective throughput rate per frame}} $\varPhi_f$, which is the ratio of requests, completed within delay requirements and in a frame, to the total backlogged and newly arrived requests before the end of that frame;
(\textit{$\romannumeral1$}) \textbf{\textit{Per frame throughput rate }} $\varPhi_f = \left[ \sum\nolimits_{n \in \mathcal{N}} \varUpsilon_{\tau}(\mathcal{Q}_{n}) + \varUpsilon_{\tau}(\mathcal{Q}_{C}) \right] / $ $\sum\nolimits_{b \in \mathcal{B}} \bar{\varUpsilon}_{\tau}(\mathcal{Q}_{b})$, which reflects the short-term characteristics of $\varPhi^{\prime}$;
(\textit{$\romannumeral2$}) \textbf{\textit{Scheduling delay}} $\varPhi_d$, the time required for a scheduling action;
(\textit{$\romannumeral3$}) \textbf{\textit{Scheduling cost}} $\varPhi_c$, primarily in terms of network bandwidth consumption, including additional packet forward due to request dispatch, and bandwidth consumption for the edge pulling service \textit{Docker} images from the cloud during service orchestration.
For clarity, we perform the necessary normalization for some metrics, and give their statistical characteristics from the results of multiple experiments.

\vspace{-0.3em}
%\subsection{Learning Ability and Scheduling Practicability}
\subsection{Learning Ability and Practicability of \textit{KaiS}}

\begin{figure}[!!!!!!!!!!!!!!hhhhhhhhhht]%[!htp]
    \centering
  \vspace{-0.6em}
    \includegraphics[width=8.85 cm]{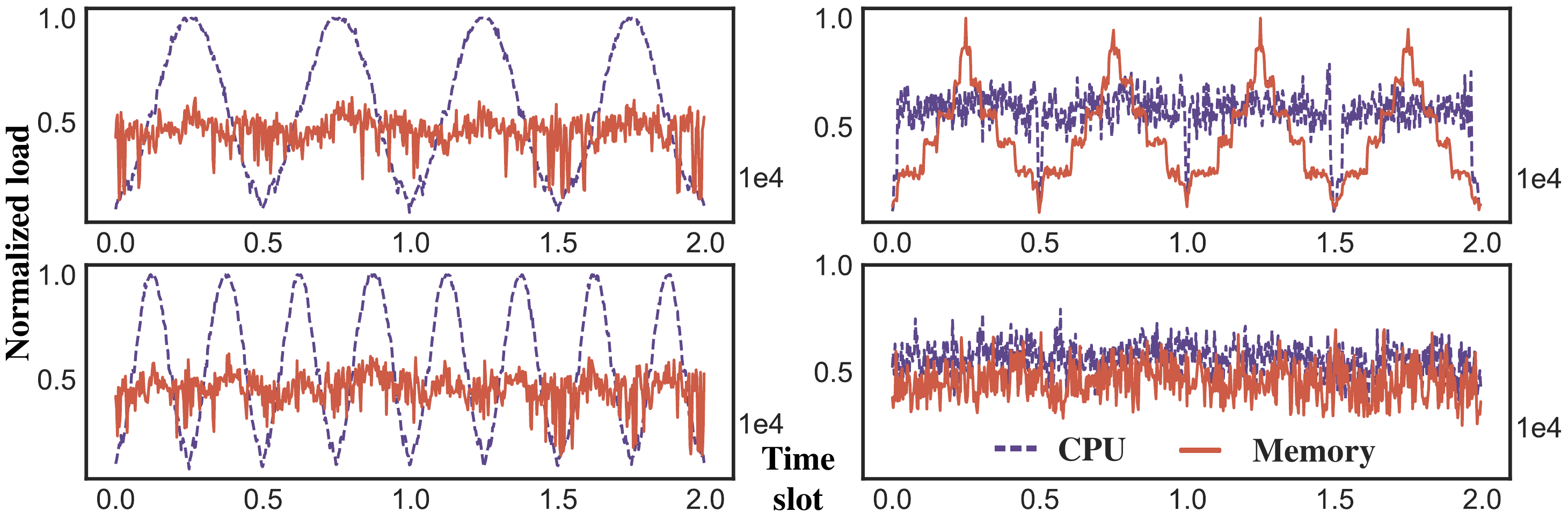}
\setlength{\abovecaptionskip}{-0.5cm} 
    \caption{Illustrative examples of 4 request arrival patterns, sampled by the service type: $\mathbb{P}_1$ (left top), $\mathbb{P}_2$ (right top), $\mathbb{P}_3$ (left bottom), $\mathbb{P}_4$ (right bottom).}
  \vspace{-0.8em}
    \label{fig:RequestPatterns}
\end{figure}

\textit{KaiS} should be able to learn how to cope with request arrivals with underlying statistical patterns and even stochastic request arrivals.
According to the service type, we sample or clip the workload dataset $\Omega$ in Sec. \ref{subsec:End-edge-cloud Computing System Setup} to acquire request arrival sequences with $4$ patterns ($20$ for each) as shown in Fig.~\ref{fig:RequestPatterns}, viz., (\textit{$\romannumeral1$}) Pattern $\mathbb{P}_1$: periodically fluctuating CPU sum load; (\textit{$\romannumeral2$}) Pattern $\mathbb{P}_2$: periodically fluctuating memory sum load; (\textit{$\romannumeral3$}) Pattern $\mathbb{P}_3$: $\mathbb{P}_1$ with $2\times$ fluctuating frequency; (\textit{$\romannumeral4$}) Pattern $\mathbb{P}_4$: raw stochastic request arrivals clipped from $\Omega$.

\begin{figure}[!!!!!!!!!!!!!!hhhhhhhhhht]%[!htp]
    \centering
%  \vspace{-0.6em}
    \includegraphics[width=8.85 cm]{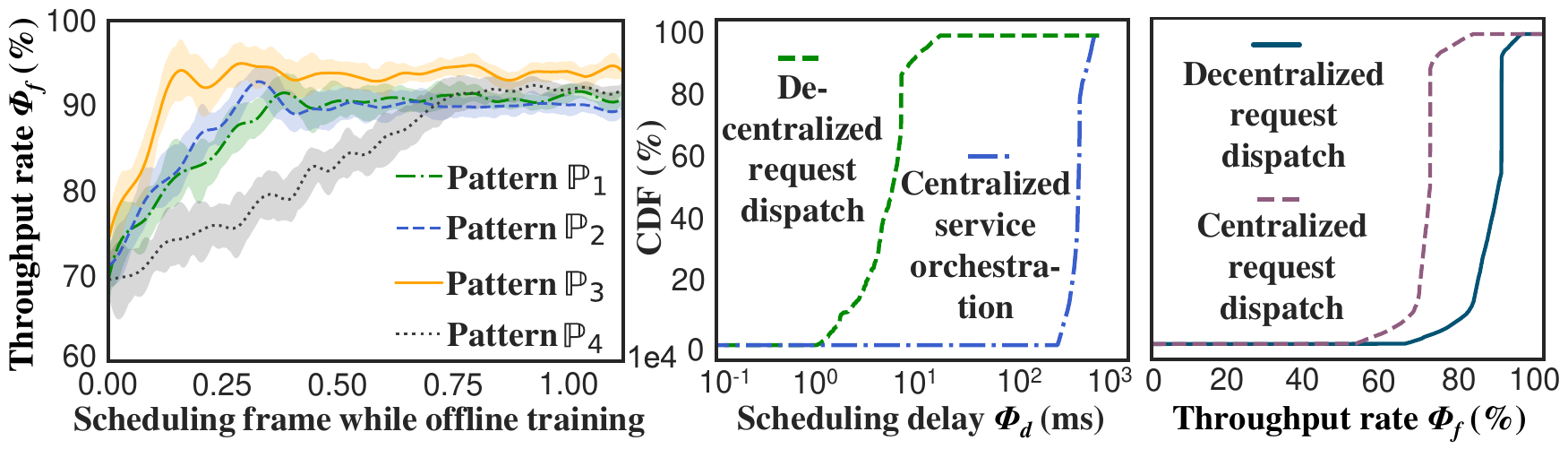}
\setlength{\abovecaptionskip}{-0.5cm} 
    \caption{(a) The learning ability of \textit{KaiS} against request arrivals with various patterns (left), and (b) the scheduling delay of \textit{KaiS} performing decentralized request dispatch and centralized service orchestration (central), and (c) the performance of \textit{KaiS} using decentralized and centralized dispatch (right).}
%  \vspace{-0.6em}
    \label{fig:LearnPatterns}
\end{figure}
\begin{figure}[!!!!!!!!!!!!!!hhhhhhhhhht]%[!htp]
    \centering
  \vspace{-0.7em}
    \includegraphics[width=8.85 cm]{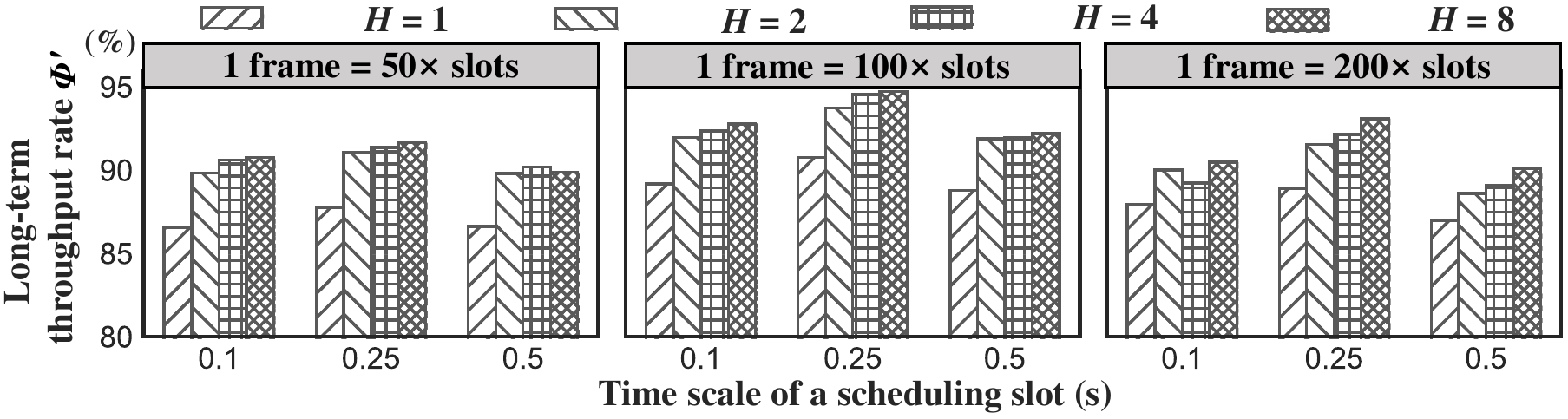}
\setlength{\abovecaptionskip}{-0.5cm} 
    \caption{\textit{KaiS}'s scheduling performance under various settings.}
  \vspace{-1.2em}
    \label{fig:HyperParameters}
\end{figure}

\begin{itemize}[leftmargin=*]
\item \textbf{\textit{Learning ability}}.
Fig.~\ref{fig:LearnPatterns}(a) gives the performance evolution of \textit{KaiS} during training for different request patterns.
The throughput rate $\varPhi_f$ in all cases is improving over time, which demonstrates that \textit{KaiS} can gradually learn to cope with different request patterns.
In particular, \textit{KaiS} requires experiencing at least $1.2$ times more frames to achieve stable scheduling, when coping with stochastic request arrivals ($\mathbb{P}_4$) rather than others ($\mathbb{P}_{1-3}$).
Nonetheless, once \textit{KaiS} converges, its scheduling performance gap for requests of different patterns is within $4.5\%$.
\item \textbf{\textit{Decentralized or centralized dispatch?}}
Fig.~\ref{fig:LearnPatterns}(b) shows that the scheduling delay of centralized service orchestration is almost $9\times$ than that of decentralized request dispatch, while the latter can be completed within around $10$~ms.
Moreover, we maintain a \textit{cMMAC} agent for each eAP in the cloud to dispatch requests in a centralized manner for comparison.
From Fig.~\ref{fig:LearnPatterns}(c), we observe that decentralized dispatch can bring higher throughput rates, since centralized dispatch requires additional time to upload local observations ($\hat{\boldsymbol{s}}_{b, t}$) and wait for dispatch decisions.
However, these extra delays are not trivial for some delay-sensitive service requests.
\item \textbf{\textit{Two-time-scale scheduling and stepwise orchestration}}. Frequent scheduling may not lead to better performance.
As shown in Fig.~\ref{fig:HyperParameters}, when a slot is $0.1$s, \textit{cMMAC} agents often experience similar system states in adjacent slots, weakening their learning abilities.
When a slot is too large ($0.5$s), the untimely dispatch also degrades performance.
Besides, too frequent service orchestration will result in more scheduling costs and make \textit{cMMAC} agents hard to converge.
Though selecting more high-value edge nodes for service orchestration at each frame can benefit the throughput, when $H \ge 2$, the improvement is very limited, while a larger $H$ leads to more scheduling cost.
The capability of \textit{KaiS} is affected by the above factors.
We will show that a default configuration ``$0.25$s (slot), $25$s (frame), $H=2$'' can already yield decent performance compared to baselines.
%The gap between the above two corroborates that decentralized dispatch is more conducive to ensure dispatch timeliness than centralized dispatch.
\end{itemize}

\begin{figure}[t]%[!htp]
    \centering
  \vspace{-0.8em}
    \includegraphics[width=8.85 cm]{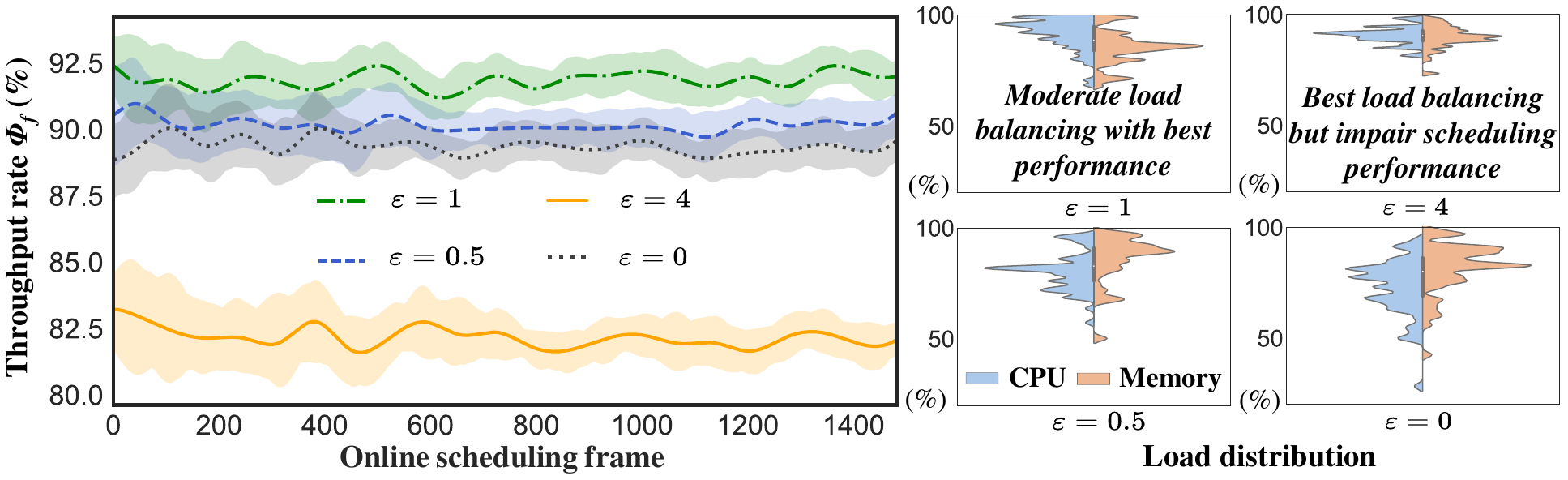}
\setlength{\abovecaptionskip}{-0.5cm} 
    \caption{Impact of load balancing on the scheduling performance of \textit{KaiS}.}
%  \vspace{-0.8em}
    \label{fig:LoadBalancing}
\end{figure}
\begin{figure}[t]%[!htp]
    \centering
  \vspace{-0.6em}
    \includegraphics[width=8.85 cm]{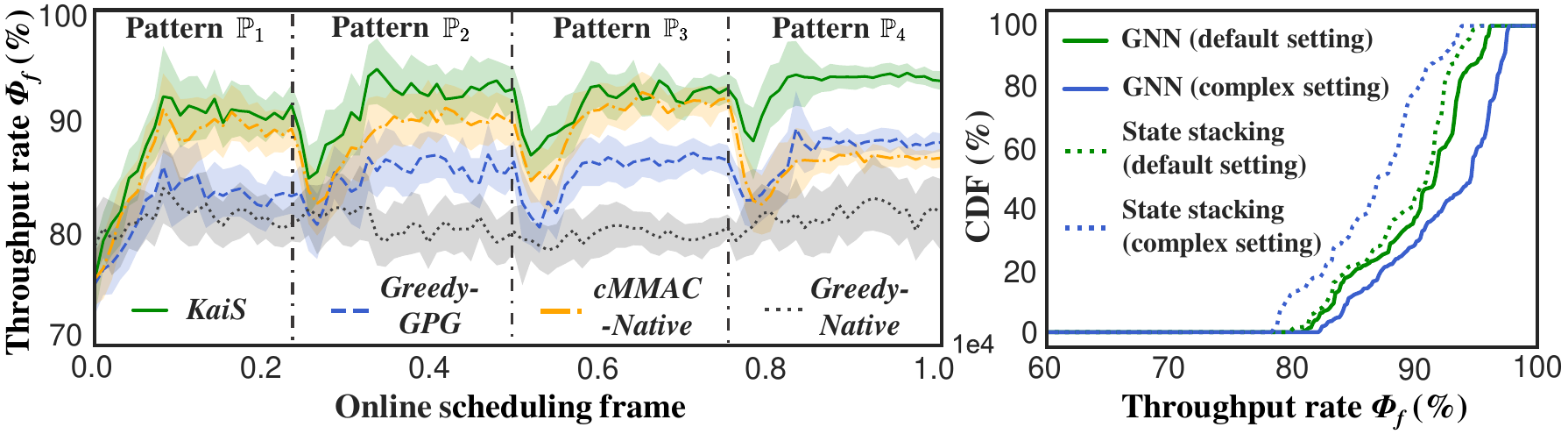}
\setlength{\abovecaptionskip}{-0.5cm} 
    \caption{The ability of \textit{GPG} to respond to (a) pattern-fluctuating request arrivals (left) and (b) different system scales and structures (right).}
  \vspace{-1.3em}
    \label{fig:GNNRole}
\end{figure}
\vspace{-0.65em}
\subsection{Impact of Load Balancing}

In Fig.~\ref{fig:LoadBalancing}, we present the scheduling performance of \textit{KaiS} trained with different settings of $\varepsilon$, which indicates the degree of edge load balancing, in $\hat{u}_{b, t} = {\text{e}}^{-\lambda - \varepsilon \nu }$.
\textit{KaiS} achieves the best throughput when $\varepsilon = 1$, while its performance sharply drops when $\varepsilon = 4$.
This performance gap lies in that, when $\varepsilon = 4$, \textit{KaiS} focuses too much on load balancing while in many cases waiving the dispatch options that can tackle requests more efficiently.
Besides, when $\varepsilon = 0$, namely load balancing is not considered, both throughput rate $\varPhi_f$ and load balancing are still better than the case $\varepsilon = 4$.
This fact demonstrates that even if we are not deliberately concerned about load balancing when designing \textit{cMMAC}, \textit{KaiS} can still learn load-balancing policies that are beneficial to improve the throughput. 
Nonetheless, setting a moderate $\varepsilon$ for the reward function can lead \textit{KaiS} to learn such policies more effectively.

\vspace{-0.3em}
\subsection{Role of GNN-based Service Orchestration}

Next, we first combine request arrival sequences of four patterns to construct a series of long one, in order to evaluate the ability of \textit{GPG} to respond to pattern-fluctuating request arrivals.
Note that these long request arrival sequences are constructed to reflect extreme scenarios with high variability.
\begin{itemize}[leftmargin=*]
\item \textbf{\textit{Coping with stochastic request arrivals}}. In Fig.~\ref{fig:GNNRole}(a), we present the scheduling performance of \textit{KaiS}, \textit{Greedy}-\textit{GPG}, \textit{cMMAC}-\textit{Native} and \textit{Greedy}-\textit{Native} to the scenarios with high variability.
The results make evident the following points: (\textit{$\romannumeral1$}) \textit{KaiS} achieves a $3.6 \%$ higher average throughput rate than the closest competing baselines, and particularly, whenever the request arrival pattern changes, \textit{KaiS} can still quickly learn a scheduling policy adapted to the new pattern; (\textit{$\romannumeral2$}) For patterns $\mathbb{P}_{1,2,3}$, \textit{cMMAC}-\textit{Native} can achieve scheduling performance close to \textit{KaiS}, the reason of which lies in that an efficient request dispatch algorithm, e.g., \textit{cMMAC}, can already address the request arrivals with obvious patterns; (\textit{$\romannumeral3$}) For the sophisticated pattern $\mathbb{P}_4$, i.e., requests are arriving stochastically as the raw traces $\Omega$, due to the lack of service orchestration to adaptively release and capture the global system resources, the performance of \textit{cMMAC}-\textit{Native} and \textit{Greedy}-\textit{Native} deteriorates.
\begin{figure}[t]%[!htp]
  \centering
%  \vspace{-0.6em}
  \includegraphics[width=8.85 cm]{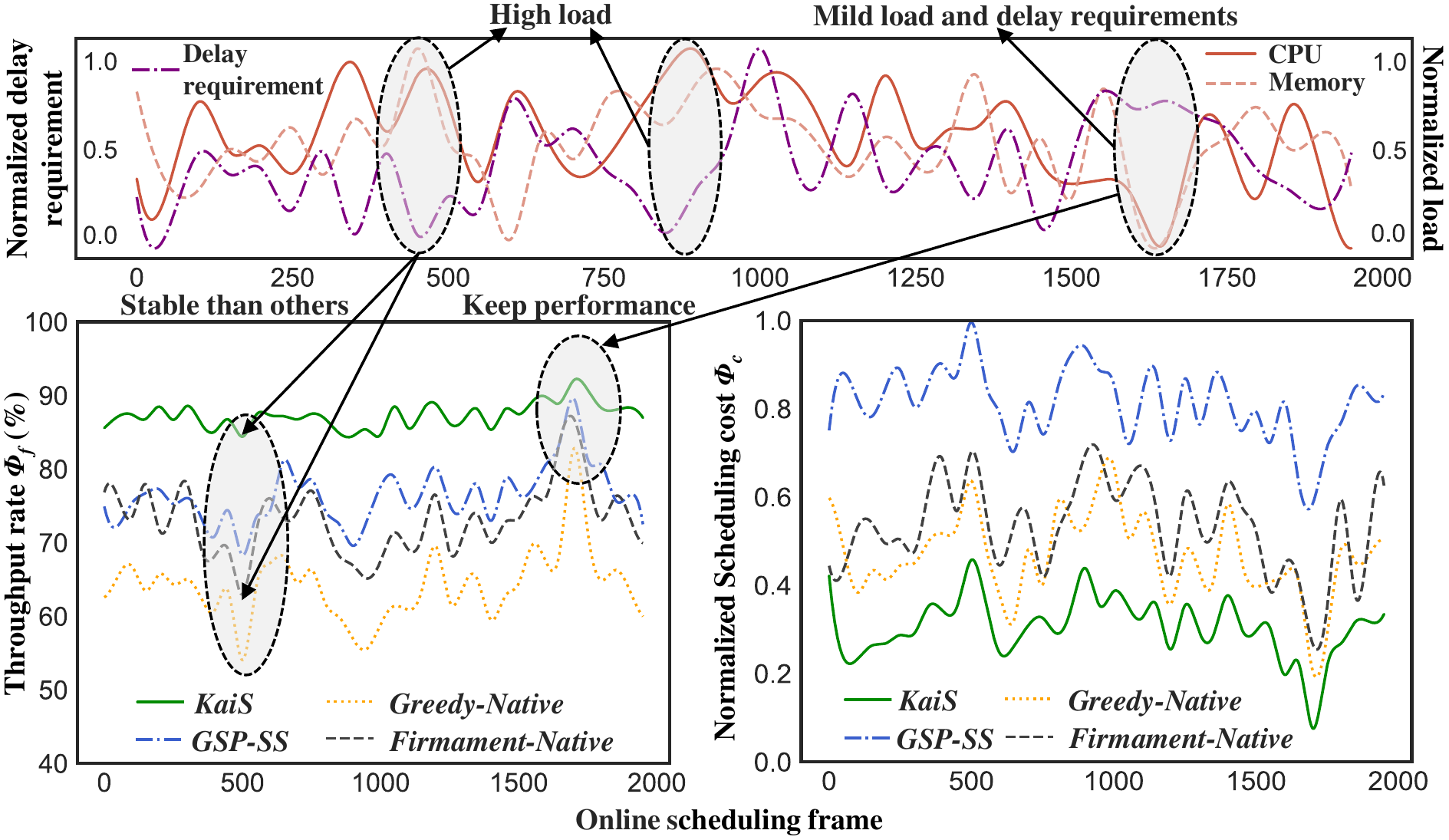}
\setlength{\abovecaptionskip}{-0.5cm} 
    \caption{(a) Under stochastic request arrivals (top), (b, c) the performance of \textit{KaiS} against baselines in terms of throughput rate (left) and cost (right).}
  \vspace{-1.2em}
  \label{fig:Baselines}
\end{figure}
%\vspace{-0.5em}
\item \textbf{\textit{GNN-based encoding against vector stacking}}. 
%For evaluating the role of \textit{GNN-based system state encoding} (Sec. \ref{subsubsec:GNN-based System State Encoding}) in \textit{KaiS}, we build two edge clusters with different scales for evaluation: (\textit{$\romannumeral1$}) a default setting introduced in Sec. \ref{subsec:End-edge-cloud Computing System Setup}; (\textit{$\romannumeral2$}) a complex setting with $10$ \textit{k3s master nodes}, each of which manages $3$-$15$ heterogeneous \textit{k3s edge nodes} ($100$ in total).
%From Fig. \ref{fig:GNNRole}(b), we observe that stacking system states into vectors cannot effectively help \textit{KaiS} fully understand the request characteristics and system information, especially in complex scenarios.
%Instead, GNN-based encoding can significantly reduce the dependence of \textit{KaiS} on the model complexity of NNs, which is key to efficient and fast learning. 
%Further, it embeds the network latency and topology information into encoded state vectors, assisting \textit{KaiS} scale to complex, large-scale edge-cloud systems.
We show in Fig.~\ref{fig:GNNRole}(b) the role of \textit{GNN-based system state encoding} (Sec. \ref{subsubsec:GNN-based System State Encoding}) in \textit{KaiS}.
For evaluation, we build two edge clusters with different system scales: (\textit{$\romannumeral1$}) a default setting introduced in Sec. \ref{subsec:End-edge-cloud Computing System Setup}; (\textit{$\romannumeral2$}) a complex setting with $10$ \textit{k3s master nodes}, each of which manages $3$-$15$ heterogeneous \textit{k3s edge nodes} ($100$ in total).
%During the comparison, other experiment settings including validating request arrival sequences and the cloud cluster remain the same. 
From Fig.~\ref{fig:GNNRole}(b), we observe that under smaller system scales, the effect of stacking system states is very close to GNN-based encoding, with only $1.3\%$ loss of scheduling performance.
However, for the complex scenario, simply stacking system states cannot effectively help \textit{KaiS} understand the request characteristics and system information, resulting in a $5.4\%$ performance loss.
Instead, GNN-based encoding can significantly reduce \textit{KaiS}' dependence on the model complexity of NNs, which is key to efficient and fast learning. 
Further, it embeds the network latency and the system structure information, assisting \textit{KaiS} scale to large-scale edge-cloud systems.
\end{itemize}

\vspace{-0.8em}
\subsection{Performance Comparison with Baselines}
\vspace{-0.2em}

To evaluate \textit{KaiS}, we need to consider both scheduling performance and cost. 
We clip the workload dataset $ \Omega $ to acquire $50$ request arrival sequences with the same length and use them to evaluate \textit{KaiS}. 
From Fig.~\ref{fig:Baselines}(b-c), we observe that in almost all cases, regardless of how the loads and the delay requirements of requests fluctuate (Fig.~\ref{fig:Baselines}(a)), \textit{KaiS} yields a $14.3 \%$ higher throughput rate $\varPhi_f$ and a $34.7 \%$ lower scheduling cost $\varPhi_c$ than the closest competing baselines.

Particularly, when the request loads and delay requirements are mild at some frames, the scheduling performance $\varPhi_f$ of \textit{GSP-SS} can be very close to that of \textit{KaiS}. 
However, in contrast to \textit{KaiS}, the scheduling performance of \textit{GSP-SS} degrades during frames with high loads: as it does not understand the system capability to process requests, when the request load level is high, it cannot load balancing the edge cluster to apportion these loads, thereby narrowing available scheduling spaces.
Besides, (\textit{$\romannumeral1$}) \textit{KaiS} adopts two-time-scale scheduling, and unlike \textit{GSP-SS} performing large-scale orchestration at each frame, (\textit{$\romannumeral2$}) it only selects a fixed number $H$ of high-value edge nodes to perform service orchestration limited by $\tilde{\mathcal{A}}$.
Hence, the scheduling cost of \textit{KaiS} is bounded in each frame, thereby reducing the overall cost, as shown in Fig.~\ref{fig:Baselines}(b). 

\vspace{-0.3em}
\section{Related Work}
\label{sec:Related Work}
\vspace{-0.2em}

Though existing optimization works explore upper bounds of scheduling performance, they are not applicable to practical deployment environments (e.g., \textit{k8s}) due to various model assumptions.
Besides, to our knowledge, there exists no system design works to accommodate decentralized request dispatch.

\subsubsection{Theoretical Analysis Work}

Many works, e.g., \cite{Zou2020,Chen2016f}, give scheduling solutions for offloading stochastic computation or service requests, which complement our work. 
The works of \cite{Farhadi2019, Poularakis2019, Ma2020} set a theoretical basis on jointly optimizing request dispatch and service orchestration.
However, the proposed one-shot scheduling optimization in \cite{Poularakis2019, Ma2020} cannot address continuously arriving service requests, i.e., without considering the long-term impact of scheduling.
In \cite{Farhadi2019}, the authors propose to perform optimization on two different time scales to maximize the number of requests completed in each schedule.
Nevertheless, the long-term optimization in \cite{Farhadi2019} relies on the accurate prediction of future service requests, which is difficult to achieve in practice.
Last but not least, these works \cite{Farhadi2019, Poularakis2019, Ma2020} cannot be applied practically since: 
(\textit{$\romannumeral1$}) They assume that the computing resources, network requirements, or the processing time for specific requests can be accurately modeled or predicted;
(\textit{$\romannumeral2$}) The dispatch is scheduled in a centralized manner, while it must take extra time to wait for the aggregation of context information across the entire system.

\subsubsection{System Design Work}

Many efficient schedulers have been developed for \textit{k8s}-based cloud clusters.
These works either schedule all tasks through minimum cost maximum flow optimization for general workloads \cite{Gog2016} or exploit domain-specific knowledge of, e.g., deep learning, to improve overall cluster utilization for specific workloads \cite{Xiao2018b}.
However, they cannot accommodate decentralized request dispatch at the edge, since their schedulers are deployed at the cloud in a centralized fashion.
The scheduler proposed in \cite{Haja2019} orchestrates services by periodically measuring the latency between edge nodes to estimate whether the expected processing delay of service requests can meet requirements.
The work most related to ours is \cite{Rossi2020}, which uses model-based RL to deal with the service orchestration and is compatible with geographically distributed edge clusters.
Nonetheless, neither \cite{Haja2019} nor \cite{Rossi2020} consider request dispatch at the edge.

\vspace{-0.3em}
\section{Conclusion}
\label{sec:Conclusion}
\vspace{-0.3em}

Leveraging \textit{k8s} to seamlessly merge the distributed edge and the cloud is the future of edge-cloud systems.
In this paper, we introduce \textit{KaiS}, a scheduling framework integrated with tailored learning algorithms for \textit{k8s}-based edge-cloud systems, that dynamically learns scheduling policies for request dispatch and service orchestration to improve the long-term system throughput rate.
Our results show the behavior of \textit{KaiS} across different scenarios and demonstrate that \textit{KaiS} can at least enhance the average system throughput rate by $14.3\%$ while reducing scheduling cost by $34.7\%$.
In addition, by modifying the scheduling action spaces and reward functions, \textit{KaiS} is also applicable to other scheduling optimization goals, such as minimizing the long-term system overhead.

\bibliographystyle{IEEEtran}
%\vspace{-0.2em}
\bibliography{INFOCOM2021}

\begin{thebibliography}{10}
\providecommand{\url}[1]{#1}
\csname url@samestyle\endcsname
\providecommand{\newblock}{\relax}
\providecommand{\bibinfo}[2]{#2}
\providecommand{\BIBentrySTDinterwordspacing}{\spaceskip=0pt\relax}
\providecommand{\BIBentryALTinterwordstretchfactor}{1}
\providecommand{\BIBentryALTinterwordspacing}{\spaceskip=\fontdimen2\font plus
\BIBentryALTinterwordstretchfactor\fontdimen3\font minus
  \fontdimen4\font\relax}
\providecommand{\BIBforeignlanguage}[2]{{%
\expandafter\ifx\csname l@#1\endcsname\relax
\typeout{** WARNING: IEEEtran.bst: No hyphenation pattern has been}%
\typeout{** loaded for the language `#1'. Using the pattern for}%
\typeout{** the default language instead.}%
\else
\language=\csname l@#1\endcsname
\fi
#2}}
\providecommand{\BIBdecl}{\relax}
\BIBdecl

\bibitem{Shi2016}
W.~Shi, J.~Cao \emph{et~al.}, ``{Edge Computing: Vision and Challenges},''
  \emph{IEEE Internet Things J.}, vol.~3, no.~5, pp. 637--646, Oct. 2016.

\bibitem{Ren2019b}
J.~Ren, D.~Zhang, S.~He, Y.~Zhang, and T.~Li, ``{A Survey on End-Edge-Cloud
  Orchestrated Network Computing Paradigms},'' \emph{ACM Comput. Surv.},
  vol.~52, no.~6, pp. 1--36, Oct. 2019.

\bibitem{Burns2016}
B.~Burns, B.~Grant, D.~Oppenheimer, E.~Brewer, and J.~Wilkes, ``{Borg, Omega,
  and Kubernetes},'' \emph{Commun. ACM}, Apr. 2016.

\bibitem{kubeedge}
``{KubeEdge}: Kubernetes native edge computing framework (project under
  {CNCF}).'' [Online]. \url{https://github.com/kubeedge/kubeedge}

\bibitem{openyurt}
``{OpenYurt}: Extending your native kubernetes to edge.'' [Online].
  \url{https://github.com/alibaba/openyurt}

\bibitem{baetyl}
``Baetyl: Extend cloud computing, data and service seamlessly to edge
  devices.'' [Online]. \url{https://github.com/baetyl/baetyl}

\bibitem{Wang2020}
X.~Wang, Y.~Han, V.~C. Leung, D.~Niyato, X.~Yan, and X.~Chen, ``{Convergence of
  Edge Computing and Deep Learning: A Comprehensive Survey},'' \emph{IEEE
  Commun. Surv. Tutor.}, vol.~22, no.~2, pp. 869--904, 2020.

\bibitem{Tan2017}
H.~Tan, Z.~Han, X.-y. Li, and F.~C. Lau, ``{Online job dispatching and
  scheduling in edge-clouds},'' in \emph{IEEE INFOCOM}, 2017.

\bibitem{Pasteris2019}
S.~Pasteris, S.~Wang, M.~Herbster, and T.~He, ``{Service Placement with
  Provable Guarantees in Heterogeneous Edge Computing Systems},'' in \emph{IEEE
  INFOCOM}, 2019.

\bibitem{Farhadi2019}
V.~Farhadi, F.~Mehmeti, T.~He, T.~L. Porta, H.~Khamfroush, S.~Wang, and K.~S.
  Chan, ``{Service Placement and Request Scheduling for Data-intensive
  Applications in Edge Clouds},'' in \emph{IEEE INFOCOM}, 2019.

\bibitem{Poularakis2019}
K.~Poularakis, J.~Llorca, A.~M. Tulino, I.~Taylor, and L.~Tassiulas, ``{Joint
  Service Placement and Request Routing in Multi-cell Mobile Edge Computing
  Networks},'' in \emph{IEEE INFOCOM}, 2019.

\bibitem{Ma2020}
X.~Ma, S.~Wang \emph{et~al.}, ``{Cooperative Service Caching and Workload
  Scheduling in Mobile Edge Computing},'' in \emph{IEEE INFOCOM}, 2020.

\bibitem{Ayala-Romero2019}
J.~A. Ayala-Romero, A.~Garcia-Saavedra, M.~Gramaglia, X.~Costa-Perez,
  A.~Banchs, and J.~J. Alcaraz, ``{vrAIn: A Deep Learning Approach Tailoring
  Computing and Radio Resources in Virtualized RANs},'' in \emph{ACM MobiCom},
  2019.

\bibitem{sutton2018reinforcement}
R.~S. Sutton \emph{et~al.}, \emph{Reinforcement learning: An introduction}.

\bibitem{Mnih2015}
V.~Mnih \emph{et~al.}, ``{Human-level control through deep reinforcement
  learning},'' \emph{Nature}, vol. 518, no. 7540, pp. 529--533, Feb. 2015.

\bibitem{lillicrap2015continuous}
T.~P. Lillicrap \emph{et~al.}, ``{Continuous control with deep reinforcement
  learning},'' \emph{arXiv preprint arXiv:1509.02971}, 2015.

\bibitem{WangMADRL}
F.~Wang, F.~Wang, J.~Liu, R.~Shea, and L.~Sun, ``{Intelligent Video Caching at
  Network Edge : A Multi-Agent Deep Reinforcement Learning Approach},'' in
  \emph{IEEE INFOCOM}, 2020.

\bibitem{MARLSurvey}
L.~{Busoniu}, R.~{Babuska}, and B.~{De Schutter}, ``A comprehensive survey of
  multiagent reinforcement learning,'' \emph{IEEE Trans. Syst., Man, Cybern. C,
  Appl. Rev.}, vol.~38, no.~2, pp. 156--172, 2008.

\bibitem{Zhang2020}
Z.~Zhang, P.~Cui, and W.~Zhu, ``{Deep Learning on Graphs: A Survey},''
  \emph{IEEE Trans. Knowl. Data Eng. (Early Access)}, 2020.

\bibitem{Silver2014}
D.~Silver, G.~Lever, N.~Heess, T.~Degris, D.~Wierstra, and M.~Riedmiller,
  ``{Deterministic policy gradient algorithms},'' in \emph{ICML}, 2014.

\bibitem{Chen2016f}
X.~Chen, L.~Jiao, W.~Li, and X.~Fu, ``{Efficient Multi-User Computation
  Offloading for Mobile-Edge Cloud Computing},'' \emph{IEEE/ACM Trans. Netw.},
  vol.~24, no.~5, pp. 2795--2808, 2016.

\bibitem{silver2016mastering}
D.~Silver \emph{et~al.}, ``Mastering the game of go with deep neural networks
  and tree search,'' \emph{Nature}, vol. 529, no. 7587, pp. 484--489, 2016.

\bibitem{greensmith2004variance}
E.~Greensmith \emph{et~al.}, ``Variance reduction techniques for gradient
  estimates in reinforcement learning,'' \emph{J. Mach. Learn. Res.}, 2004.

\bibitem{k3s}
``Lightweight kubernetes.'' [Online]. \url{https://github.com/rancher/k3s}

\bibitem{AlibabaDatasets}
``Aliababa-clusterdata.'' [Online].
  \url{https://github.com/alibaba/clusterdata}

\bibitem{k8sNativeScheduler}
``{K8s documentation: horizontal pod autoscaler}.'' [Online].
  \url{https://kubernetes.io/docs/tasks/run-application/horizontal-pod-autoscale/}

\bibitem{Gog2016}
I.~Gog, M.~Schwarzkop \emph{et~al.}, ``{Firmament: Fast, centralized cluster
  scheduling at scale},'' in \emph{USENIX OSDI}, 2016.

\bibitem{Zou2020}
J.~Zou, T.~Hao, C.~Yu, and H.~Jin, ``{A3C-DO: A Regional Resource Scheduling
  Framework based on Deep Reinforcement Learning in Edge Scenario},''
  \emph{IEEE Trans. Comput. (Early Access)}, 2020.

\bibitem{Xiao2018b}
W.~Xiao, Z.~Han \emph{et~al.}, ``{Gandiva: Introspective cluster scheduling for
  deep learning},'' in \emph{USENIX OSDI}, 2018.

\bibitem{Haja2019}
D.~Haja, M.~Szalay, B.~Sonkoly, G.~Pongracz, and L.~Toka, ``{Sharpening
  Kubernetes for the Edge},'' in \emph{ACM SIGCOMM Posters and Demos}, 2019.

\bibitem{Rossi2020}
F.~Rossi, V.~Cardellini, F.~{Lo Presti}, and M.~Nardelli, ``{Geo-distributed
  efficient deployment of containers with Kubernetes},'' \emph{Comput.
  Commun.}, vol. 159, pp. 161--174, Jun. 2020.

\end{thebibliography}

\end{document}